\documentclass[9pt,shortpaper,twoside,web]{ieeecolor}
\usepackage{generic}
\usepackage{cite}
\usepackage{url} 
\usepackage{booktabs}
\usepackage{epstopdf}
\usepackage{float}
\usepackage{stfloats}
\usepackage{subfigure}
\usepackage{multirow}
\usepackage[table,xcdraw]{xcolor}
\usepackage{multirow}
\usepackage{amsmath,amssymb,amsfonts}
\usepackage{algorithmic}
\usepackage{algorithm} 
\usepackage{graphicx}
\usepackage{textcomp}
\usepackage{threeparttable}
\def\BibTeX{{\rm B\kern-.05em{\sc i\kern-.025em b}\kern-.08em
    T\kern-.1667em\lower.7ex\hbox{E}\kern-.125emX}}
\begin{document}
\title{Deep Learning Methods for Lung Cancer Segmentation in Whole-slide Histopathology Images - the ACDC@LungHP Challenge 2019}
\author{ Zhang Li, Jiehua Zhang, Tao Tan, Xichao Teng, Xiaoliang Sun, Yang Li, Lihong Liu, Yang Xiao,  Byungjae Lee, Yilong Li, Qianni Zhang, Shujiao Sun, Yushan Zheng, Junyu Yan, Ni Li, Yiyu Hong, Junsu Ko, Hyun Jung, Yanling Liu, Yu-cheng Chen, Ching-wei Wang, Vladimir Yurovskiy, Pavel Maevskikh, Vahid Khanagha, Yi Jiang, Xiangjun Feng, Zhihong Liu, Daiqiang Li, Peter J. Sch{\"u}ffler, Qifeng Yu, Hui Chen, Yuling Tang, Geert Litjens 
\thanks{This work was partially suppored by National Natural Science Funding of China (No.61801491) and Natural Science Funding of Hunan Province (No.2019JJ50728). }
\thanks{J. Zhang, Z. Li, X. Teng, X. Sun, Q. Yu and Y. Li are with College of Aerospace Science and Engineering, National University of Defense Technology, Changsha 410073, China}
\thanks{T. Tan is with Department of Mathematics and Computer Science, Eindhoven University of Technology, Eindhoven 5600 MB, The Netherlands and ScreenPoint Medical, Nijmegen, The Netherlands}
\thanks{H. Liu and Y. Xiao are with Pingan Technology, Shenzhen, China}
\thanks{B. Lee is with Lunit Inc., Seoul, Korea}
\thanks{Y. Li, and Q. Zhang are with School of Electrical Engineering
and Computer Science, Queen Mary University of London,
London, United Kingdom}
\thanks{S. Sun is with Image Processing Center, School of Astronautics, Beihang University, Beijing 102206, China and Beijing Advanced Innovation Center for Biomedical Engineering, Beijing 100191, China }
\thanks{Y. Zheng is with Beijing Advanced Innovation Center for Biomedical Engineering, Beihang University and Image Processing Center, School of Astronautics, Beihang University, Beijing 102206, China.}
\thanks{J. Yan and N. L are with  AstLab, School of Automatationm, Beihang University, Beijing, China}
\thanks{Y. Hong and J. Ko are with Department of R\&D Center, Arontier Co., Ltd., Seoul, Korea}
\thanks{H. Jung and Y. Liu are with Frederick National Laboratory}
\thanks{Y. Chen and C. Wang are with the Center of Computer Vision and Medical Imaging, the Graduate Institute of Biomedical Engineering and the Graduate institute of Applied Science and Technology, National Taiwan University of Science and Technology and AI Explore, Taipei, Taiwan}
\thanks{P. Maevskikh and V. Yurovskiy are with Research Department Skychain Global, Yekaterinburg, Russia}
\thanks{V. Khanagha is with Audio Solutions Team, Motorola Solutions Inc.
Plantation, Florida, USA}
\thanks{Y. Jiang and D. Li with The Second Xiangya Hospital, Central South University, Changsha, China}
\thanks{X. Feng is with Hunan Biotechnology Ltd., Changsha, China}
\thanks{Z. Liu is with Hunan Cancer Hospital, Central South University, Changsha, China}
\thanks{P. J. Sch{\"u}ffler is with Memorial Sloan Kettering Cancer Center, USA}
\thanks{H. Chen, and Y. Tang are with The First Hospital of Changsha City, Changsha, China}
\thanks{G. Litjens is with Radboud University Medical Center in Nijmegen, the Netherlands}
\thanks{*Zhang Li, Jiehua Zhang and Tao Tan have equal contribution to this work }
\thanks{**Corresponding author: Zhang Li (email:zhangli\_nudt@163.com); 
     Hui Chen(email:achma8088@163,com);     
     Yuling Tang(email:tyl71523@sina.com)}
}
\maketitle
\begin{abstract}
Accurate segmentation of lung cancer in pathology slides is a critical step in improving patient care. We proposed the ACDC@LungHP (Automatic Cancer Detection and Classification in Whole-slide Lung Histopathology) challenge for evaluating different computer-aided diagnosis (CADs) methods on the automatic diagnosis of lung cancer. The ACDC@LungHP 2019 focused on segmentation (pixel-wise detection) of cancer tissue in whole slide imaging (WSI), using an annotated dataset of 150 training images and 50 test images from 200 patients.
This paper reviews this challenge and summarizes the top 10 submitted methods for lung cancer segmentation. All methods were evaluated using the false positive rate, false negative rate, and DICE coefficient (DC). The DC ranged from 0.7354$\pm$0.1149 to 0.8372$\pm$0.0858. The DC of the best method was close to the inter-observer agreement (0.8398$\pm$0.0890). All methods were based on deep learning and categorized into two groups: multi-model method and single model method. In general, multi-model methods were significantly better (\textit{p}$<$0.01) than single model methods, with mean DC of 0.7966 and 0.7544, respectively.
Deep learning based methods could potentially help pathologists find suspicious regions for further analysis of lung cancer in WSI.
\end{abstract}


\section{INTRODUCTION}
\label{sec:introduction}
Lung cancer is the top cause of cancer-related death in the world. According to the 2009-2013 SEER (Surveillance, Epidemiology, and End Results) database, the 5-year survival rate of lung cancer patients is approximately 18\% \cite{1}. For patients with the early stage, resectable cancer, the 5-year survival rate is about 34\%, but for unresectable cancer, the 5-year survival rate is less than 10\%. Therefore, early detection and diagnosis of lung cancer are the key important steps in improving patient treatment outcomes. According to the National Comprehensive Cancer Network (NCCN) guidelines, for image-suspected tumors, histopathological assessment of biopsies obtained via fiberoptic bronchoscopy should be performed for the diagnosis \cite{2,3}.

Assessment of biopsy tissue by a pathologist is the golden standard for lung cancer diagnosis. However, the diagnostic accuracy was less than 80\% \cite{4}. The major histological subtypes of malignant lung disease are squamous carcinoma, adenocarcinoma, small cell carcinoma, and undifferentiated carcinoma. Correctly assessing these subtypes on biopsy is paramount for correct treatment decisions. However, the number of qualified pathologists is too small to meet the substantial clinical demands, especially in countries such as China, with a significant population of lung cancer patients. Recently, the results from the largest randomized control lung screening trial, the National Lung Screening Trial (NLST), led to the implementation of lung cancer screening with low-dose Computed Tomography in the United States in 2015. Moreover, the results from the second-largest randomized control trial, the Dutch-Belgian lung cancer screening trial (NELSON), also show the benefits of implementing lung cancer screening. The implementation in the U.S. and the possible implementation of lung cancer screening in Europe will likely lead to a substantial amount of whole-slide histopathology images biopsies and resected tumors. At the same time, the workload and the shortage of pathologists are severe. An artificial intelligence (AI) system might efficiently solve the problems mentioned above by an automatic assessment of lung biopsies.

Digital pathology has been gradually introduced in pathological clinical practice. Digital pathology scanners could generate high-resolution WSIs (up to 160nm per pixel). It facilitates the development of automatic analysis algorithms for reducing the burden and improving the performance of pathologists. Most recently, a large number of deep learning (DL) methods have been proposed for automatic image analysis of WSIs from the cell level to the image level\cite{5,6,7,8,9,10,11} .

At the cell level, DL methods were used in mitosis detection \cite{12,13,14}, nucleus detection \cite{15,16,17} and cell classification\cite{18,19}. These proposed DL methods were all based on convolutional neural networks (CNNs). At the tissue level, several CNNs were also proposed for tissue segmentation (e.g., segmenting glands for grading adenocarcinomas \cite{20}). Moreover, contour information\cite{21}, handcrafted features \cite{22,23}, multi-loss \cite{24,25,26} were incorporated into CNNs to obtain reliable tissue segmentation results.

At the image level, a three-layers CNN was first introduced to detect invasive ductal breast carcinoma and showed a comparable result (65.40\% accuracy) with classifiers relying on specific handcrafted features \cite{27}. CNNs were also used in the detection of prostate cancer\cite{28}, pancreas cancer \cite{29} and kidney cancer \cite{30}. Furthermore, CNN could be used as a features extractor for colon cancer classification and colon cancer prediction \cite{31}. 

Deeper CNN, such as GoogLeNet \cite{32}, AlexNet \cite{33}, VGG \cite{34} and ResNet \cite{35}, was transferred to breast cancer classification \cite{36} and prostate cancer prediction \cite{37}. In the CAMELYON16 challenge \cite{38}, the 1st rank team ensembled two GoogLeNets to elevate the AUC of classification lymph node metastases to 99.4\%. Several challenges in medical imaging also significantly advanced the pathology image analysis community, such as mitosis detection challenges in ICPR 2012\footnote{\url{http://people.idsia.ch/~juergen/deeplearningwinsMICCAIgrandchallenge.html}},     CAMELYON16\footnote{\url{https://camelyon16.grand-challenge.org/}}  and CAMELYON17\footnote{\url{https://camelyon17.grand-challenge.org/}}  for identifying breast cancer metastases. In particular, the CAMELYON16 was the first challenge to offer WSIs a large number of annotations, which is essential for training larger CNNs such as ResNet.

With the breakthrough of DL methods in medical image analysis and increasing of available public WSIs for developing a specific CNN, we believe that the CNN could be leveraged to give pathologists more reliable objective results or even help pathologists to improve the cancer diagnostic level. However, after assessing recent review papers \cite{10,11}, we found very few articles discussing the applications of CNNs to histopathological images of lung cancer. Furthermore, no public datasets of WSI were available to evaluate such algorithms. A recent paper that used CNNs on lung cancer detection was only on cytological image \cite{39}. The size of each image was limited (only around 1k*1k pixels), and the appearance of this image was quite different from the hematoxylin\&eosin (H\&E) stained image that we used in this paper. The recent research \cite{40} suggested that image features automatically extracted from WSIs can predict the prognosis of lung cancer patients and thereby contribute to precision oncology by machine learning classifiers. 

\begin{figure}[htpb]
\centerline{\includegraphics[width=5cm,height=5cm]{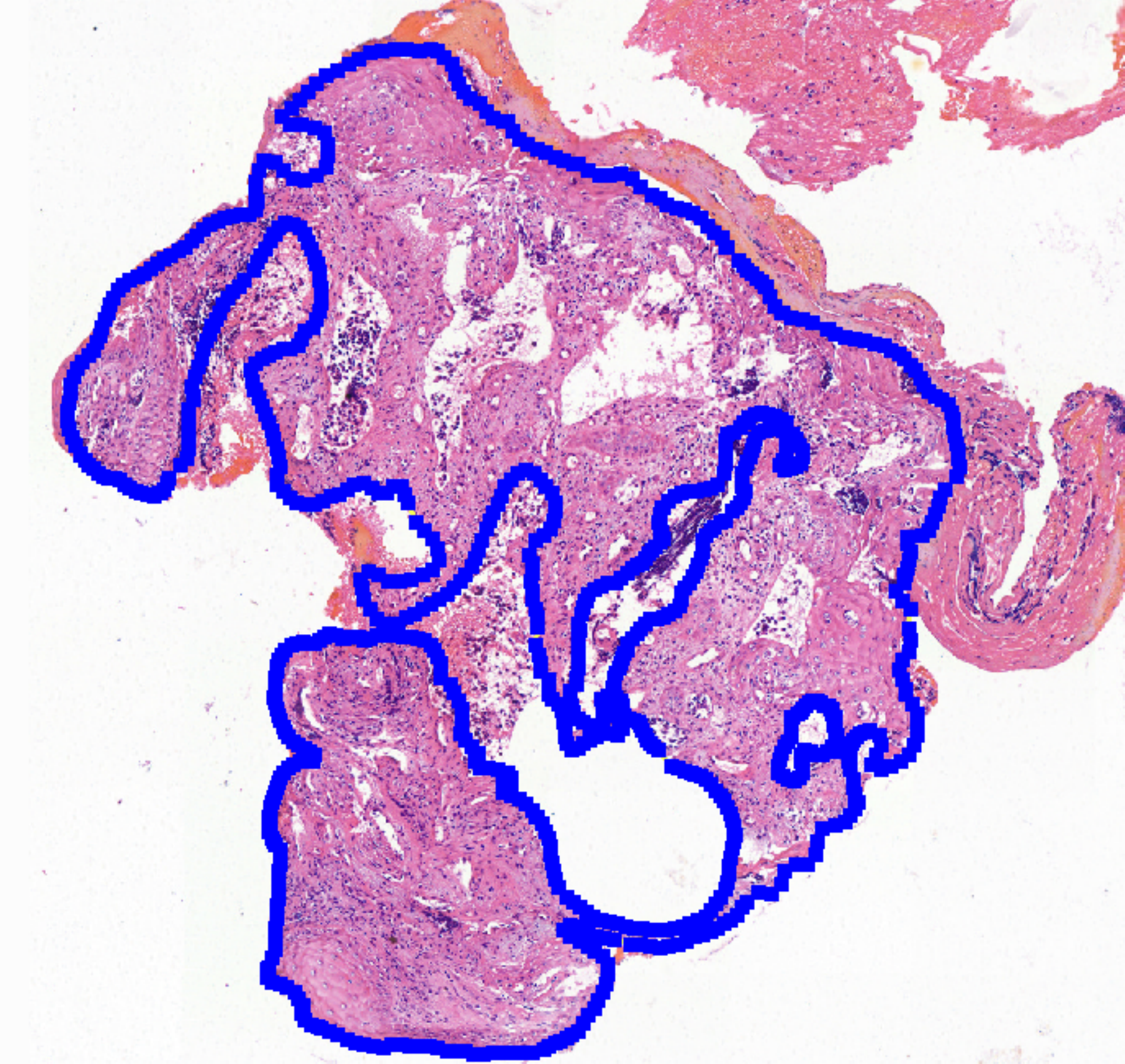}}
\caption{Pathological WSI with annotations for cancer regions.}
\label{Fig 1}
\end{figure}

To further explore the potential application of DL on WSI for lung cancer diagnosis, we proposed the ACDC@LungHP challenge which is the first one addressing lung cancer detection and classification using WSI, to our best knowledge \cite{36}. This manuscript is a summary of the first stage of ACDC@LungHP (in conjunction with ISBI2019) that focused on the segmentation of cancer tissue in WSI. The sample of pathological WSI with annotations was shown in Fig.\ref{Fig 1}. 

\section{Materials}
\label{sec:data}
\subsection{Patient recruitment}
For the ACDC@LUNGHP challenge, 200 lung patients were recruited in this study at the Department of Pulmonary Oncology in the First Hospital of Changsha, from January 2016 to November 2017. According to the American Joint Committee on Cancer (AJCC) staging system, patients firstly diagnosed with lung/bronchus cancer (site: C34.1-C34.9; histology type: adenocarcinoma, squamous cell carcinoma, and small cell carcinoma) were recruited. Other inclusion criteria included: 1) pathologically confirmed patients with surgery biopsy maintained; 2) no radiotherapy before surgery; 3) aged between 30 and 90 yr. The exclusion criteria were: 1) multiple primary cancers; 2) metastatic lung cancer; 3) patients with immune-deficiency or organ-transplantation history; 4) patients who did not provide informed consent. This study was approved by the Ethics Committee of the First Hospital of Changsha. Informed consent was obtained from each patient before the examination. Necessary demographic and clinical information for each patient, such as age, gender, stage, pathology, etc. were collected.
\subsection{Data preparation}
\begin{figure}[htpb]
\centerline{\includegraphics[width=\columnwidth]{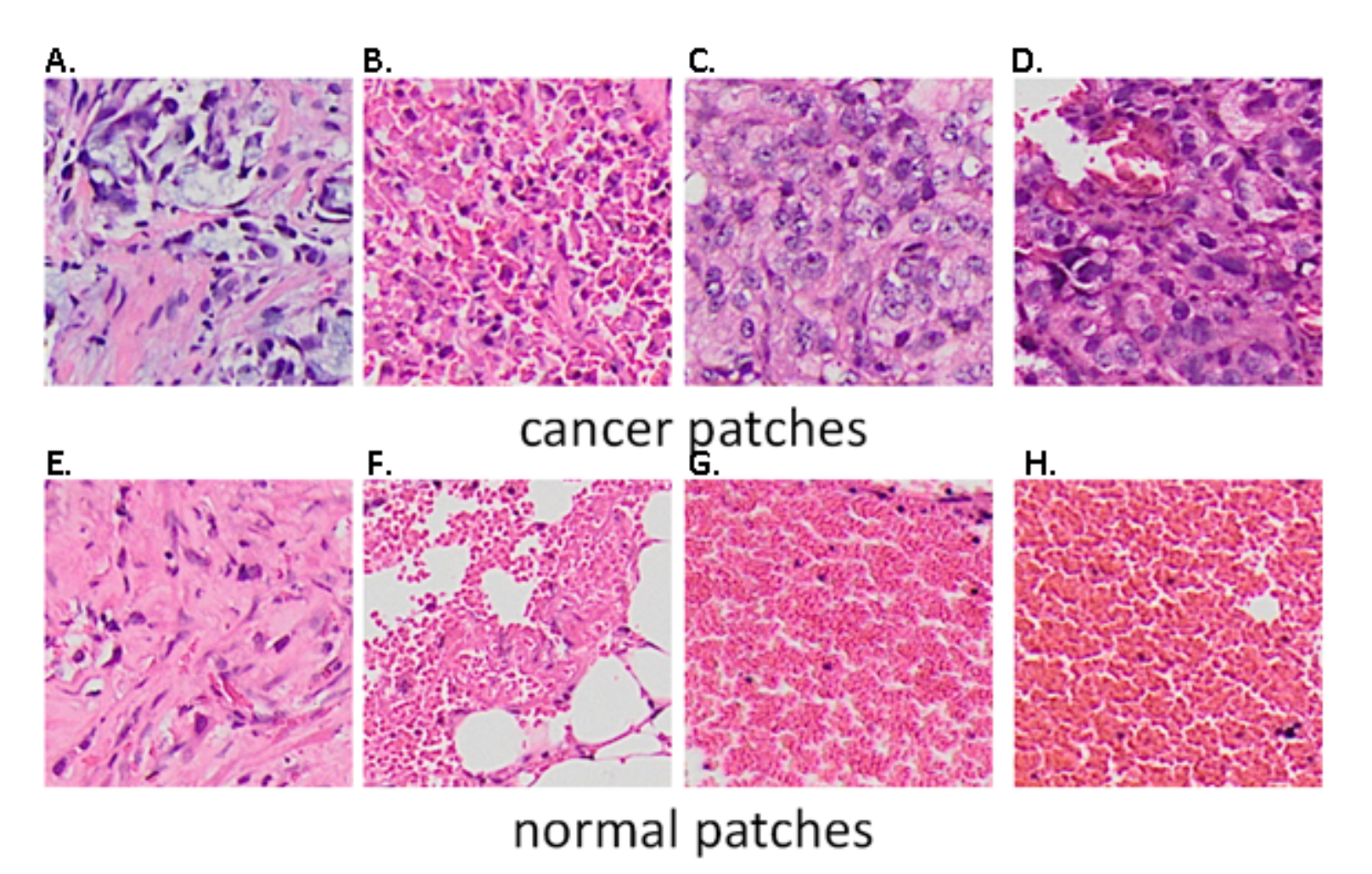}}
\caption{ Example of tumor patches and normal patches.}
\label{Fig 2}
\end{figure}

Histological slides were stained with H\&E scanned by a digital slide scanner (3DHISTECH Pannoramic 250) with objective magnifications of 20x. The close look of different tissues in the slides can be seen in Fig.\ref{Fig 2}. One can see that the patch colors were quite different even among the patches from normal tissue due to the staining variability. The appearance of the cancer regions was also quite different because of the different cancer types. For instance, Fig.\ref{Fig 2}.(A) and (B) represented small-cell lung cancer, and Fig.\ref{Fig 2}.(C) and (D) represented squamous cell lung cancer and adenoid cell lung cancer. Fig.\ref{Fig 2}.(E)-(H) were normal patches. 

In total, 200 H\&E stained slides were scanned and digitized. We randomly split those 200 slides into training and test sets. 150 slides with annotation were released as the training set. 50 slides were held as the test set. The main types of cancer were included in our data: squamous cell carcinoma, small cell carcinoma, and adenocarcinoma. The ratio of them was approximately 6:3:1. One pathologist with 30 years of experience (the director of the pathology department) annotated the cancer regions for all 200 slides (See Fig.\ref{Fig 1}). We also asked the second pathologist (with 20 years experience) to annotate the test set only. The annotation of the second pathologist was only used for accessing the inter-observer variability. Participants were allowed to use their own training data for pre-training. All data were uploaded to Microsoft OneDrive, Google Drive, and Baidu Pan for participants from different regions. Whole-Slide images were released in the TIFF format. Manual annotations were in XML format.

In the clinical practice, more than one sample from the same biopsy will be scanned. If samples have a similar shape, the pathologist only annotated one sample for the WSI. Participants were suggested to use ASAP\footnote{\url{https://github.com/GeertLitjens/ASAP}} to make a bounding box themselves to exclude the unused samples. 

\section{The ACDC@LUNGHP challenge summary}

\subsection{Challenge overview}
\begin{figure}[htpb]
\centerline{\includegraphics[width=4.5cm]{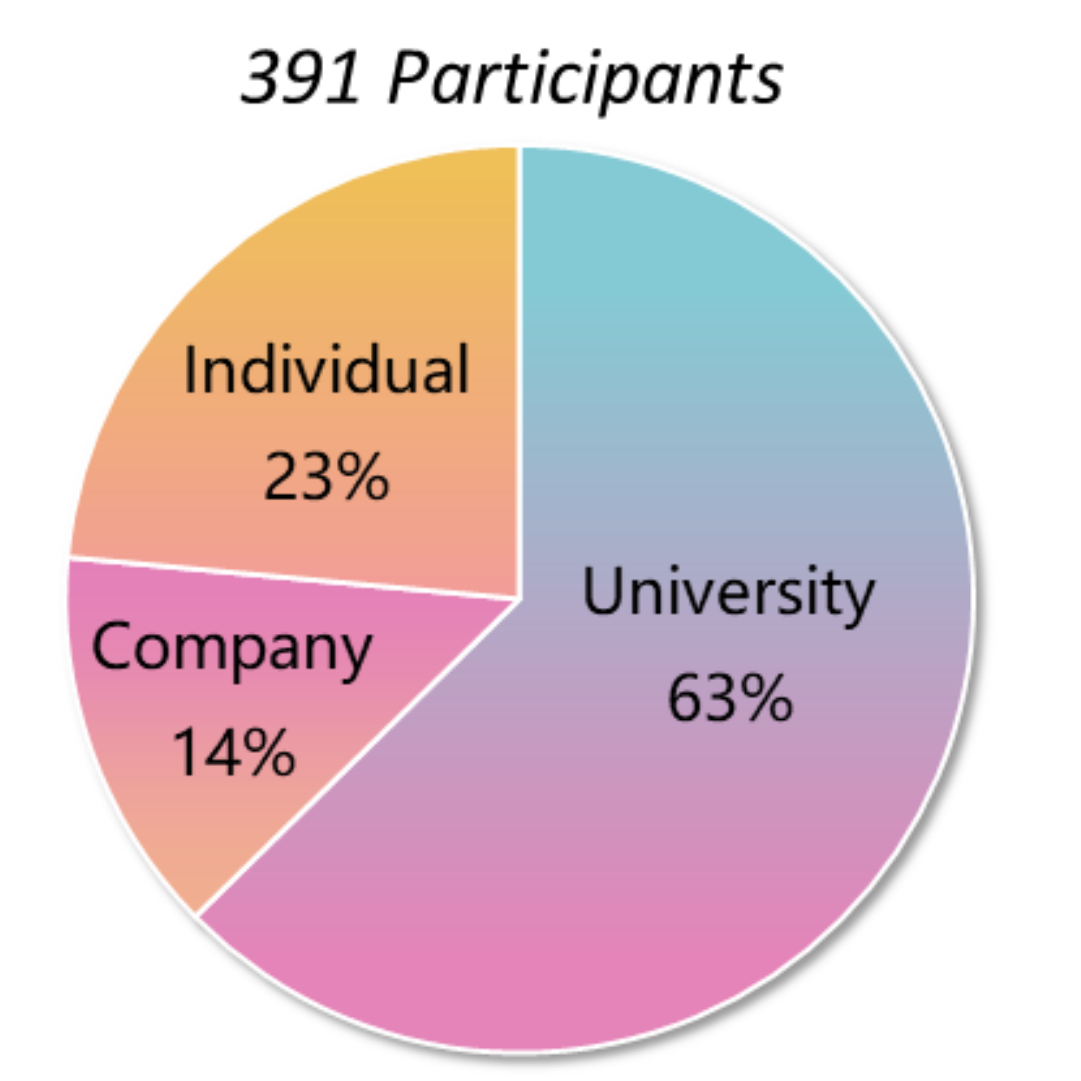}}
\caption{ The distribution of participants from ACDC@LungHP 2019.}
\label{Fig 3}
\end{figure}

The first stage of the ACDC@LUNGHP challenge focused on detecting and segmenting lung carcinoma in WSI. The segmentation as a potential aid could quickly help pathologists to identify suspicious regions. At this stage, 495 participants submitted the challenge applications, and 391 of them were confirmed as valid participants (with required registration information). Each team was allowed to submit their result three times per day. 25 participants successfully submitted their results before the closing time. The distribution of the participants was shown in Fig.\ref{Fig 3}.

The Dice coefficient (DC) was computed to evaluate the agreement between the automatic segmentation and the manual annotation by the pathologist. The DC was defined as:
\begin{equation}
\emph{Dice}=\frac{2|GT\cap{RES}|}{|GT|+|RES|}
\end{equation}
where the GT and RES are ground truth from the pathologist and result of automatic segmentation. The top 10 teams were selected from the final participants. The overall comparison could be seen in Table \ref{tab1}. The DC ranges from 0.7354 to 0.8372. Based on the model ensembling strategy (see the following sections), the methods from top 10 teams could be categorized into two groups: multi-model method and single model method. Other criteria, such as label refine, pre-processing and pre-training strategy were also summarized in Table \ref{tab1}.   
\begin{table*}[ht]
\centering
 \caption{overall comparison of top 10 teams
			 for the ACDC@LUNGHP challenge 2019.}\label{tab1}
\begin{tabular}{|c|l|l|c|c|l|l|l|l|l|l|l|l|l|l|l|}
\hline
\multicolumn{1}{|l|}{\multirow{2}{*}{}}                                             & \multicolumn{1}{c|}{\multirow{2}{*}{\textbf{Team}}} & \multicolumn{2}{l|}{\textbf{\begin{tabular}[c]{@{}l@{}}Task:\\ Lung cancer\\ detection\end{tabular}}} & \multicolumn{12}{c|}{\textbf{Deep Learning Algorithm}}                                                                                                                                                                                                                                                                                                                                                                                                                                                          \\ \cline{3-16} 
\multicolumn{1}{|l|}{}                                                              & \multicolumn{1}{c|}{}                               & \multicolumn{1}{c|}{\textbf{\begin{tabular}[c]{@{}c@{}}Mean\\ DC\end{tabular}}}   & \textbf{Rank}   & \multicolumn{2}{c|}{\textbf{\begin{tabular}[c]{@{}c@{}}Label\\ refine\end{tabular}}} & \multicolumn{1}{c|}{\textbf{Architecture}}                                        & \multicolumn{1}{c|}{\textbf{Preprocessing}}                                                                                                            & \multicolumn{8}{c|}{\textbf{Comments}}                                                                                                                                      \\ \hline
\multirow{5}{*}{\textbf{\begin{tabular}[c]{@{}c@{}}\\ \\ \\ \\Multi\\ Model\end{tabular}}}     & \textbf{PATECH}                                     & 0.8372                                                                              & 1               & \multicolumn{2}{c|}{$\surd$}                                                               & \begin{tabular}[c]{@{}l@{}}DenseNets\&\\ dilation block\\ with U-Net\end{tabular} & \begin{tabular}[c]{@{}l@{}}Color normalization; Ostu \\ to refine label\end{tabular}                                                                   & \multicolumn{8}{l|}{\begin{tabular}[c]{@{}l@{}}Capture more context information \\ and multi-scale feature;\\ Ensemble of models by changing \\ loss function\end{tabular}} \\ \cline{2-16} 
                                                                                    & \textbf{Byungjae Lee}                               & 0.8297                                                                              & 2               &  \multicolumn{2}{c|}{$\surd$}                                                               & \begin{tabular}[c]{@{}l@{}}ResNet50\&\\ DeepLab V3+\end{tabular}                  & \begin{tabular}[c]{@{}l@{}}Multi data augmentations;\\ Ostu to refine label\end{tabular}                                                               & \multicolumn{8}{l|}{\begin{tabular}[c]{@{}l@{}}Initialize encoder with ImageNet \\ pre-trained weights\end{tabular}}                                                        \\ \cline{2-16} 
                                                                                    & \textbf{Turbolag}                                   & 0.7968                                                                              & 3               & \multicolumn{2}{c|}{}                                                                & \begin{tabular}[c]{@{}l@{}}U-Net\&\\ ConvCRF\end{tabular}                         & \begin{tabular}[c]{@{}l@{}}Multi-resolution training \\ data\end{tabular}                                                                              & \multicolumn{8}{l|}{\begin{tabular}[c]{@{}l@{}}Multiple networks; enhance the \\ boundary accuracy\end{tabular}}                                                            \\ \cline{2-16} 
                                                                                    & \textbf{ArontierHYY}                                & 0.7638                                                                              & 6               & \multicolumn{2}{c|}{$\surd$}                                                               & \begin{tabular}[c]{@{}l@{}}Mdrn80+DenseNet\&\\ ResNet\end{tabular}                & Tile labeling strategy                                                                                                                                 & \multicolumn{8}{l|}{Ensemble of 16 models}                                                                                                                                  \\ \cline{2-16} 
                                                                                    & \textbf{Newhyun00}                                  & 0.7552                                                                              & 7               & \multicolumn{2}{c|}{$\surd$}                                                               & DenseNet103                                                                       & Select clean labels                                                                                                                                    & \multicolumn{8}{l|}{\begin{tabular}[c]{@{}l@{}}”Co-teaching” method made \\ training deep neural networks \\ robustly\end{tabular}}                                         \\ \hline
\multirow{5}{*}{\textbf{\begin{tabular}[c]{@{}c@{}}\\ \\ \\ \\ \\ \\Single\\  Model\end{tabular}}} & \textbf{CMIAS}                                      & 0.7700                                                                              & 4               & \multicolumn{2}{c|}{$\surd$}                                                               & \begin{tabular}[c]{@{}l@{}}DenseNet121\&\\ FCN\end{tabular}                       & \begin{tabular}[c]{@{}l@{}}Locate the tissue regions\\ by a bounding box\end{tabular}                                                                  & \multicolumn{8}{l|}{Combination of two networks}                                                                                                                            \\ \cline{2-16} 
                                                                                    & \textbf{Jorey}                                      & 0.7659                                                                              & 5               & \multicolumn{2}{c|}{$\surd$}                                                               & \begin{tabular}[c]{@{}l@{}}IncRes+ACF\&\\ CRF\end{tabular}                        & \begin{tabular}[c]{@{}l@{}}Ostu to refine label; \\ Divided into 3 classes\\ (tumor; normal;mix)\\ and mix Mix file into \\ other classes\end{tabular} & \multicolumn{8}{l|}{\begin{tabular}[c]{@{}l@{}}Feature fusing by using\\ multi-atrous convolution\end{tabular}}                                                             \\ \cline{2-16} 
                                                                                    & \textbf{AIExplore}                                  & 0.7510                                                                              & 8               & \multicolumn{2}{c|}{}                                                                & FCN                                                                               & None                                                                                                                                                   & \multicolumn{8}{l|}{\begin{tabular}[c]{@{}l@{}}Training in the AI Explore\\ platform; Using a large\\ momentum in SGD; Pre-trained \\ network\end{tabular}}                 \\ \cline{2-16} 
                                                                                    & \textbf{Skyuser}                                    & 0.7456                                                                              & 9               & \multicolumn{2}{c|}{}                                                                & ResNet18                                                                          & Multi data augmentations;                                                                                                                              & \multicolumn{8}{l|}{\begin{tabular}[c]{@{}l@{}}Classifier-based approach; \\ Fast, small and robust network\end{tabular}}                                                   \\ \cline{2-16} 
                                                                                    & \textbf{Vahid}                                      & 0.7354                                                                              & 10              & \multicolumn{2}{c|}{}                                                               & Small-FCN-512                                                                     & None                                                                                                                                                   & \multicolumn{8}{l|}{\begin{tabular}[c]{@{}l@{}}Designed a custom FCN; Pre-trained \\ network\end{tabular}}                                                                  \\ \hline
\end{tabular}
\end{table*}

\subsection{The methods based on single model}
The single model methods only used the individual model as their architectures. The mean DC for single model methods was 0.7544. An overall comparisons of single model methods could be seen in Table \ref{tab1}). 

\begin{figure*} \centering    
\subfigure[patch classification based on CNN.] {
 \label{fig:a}     
\includegraphics[width=12cm]{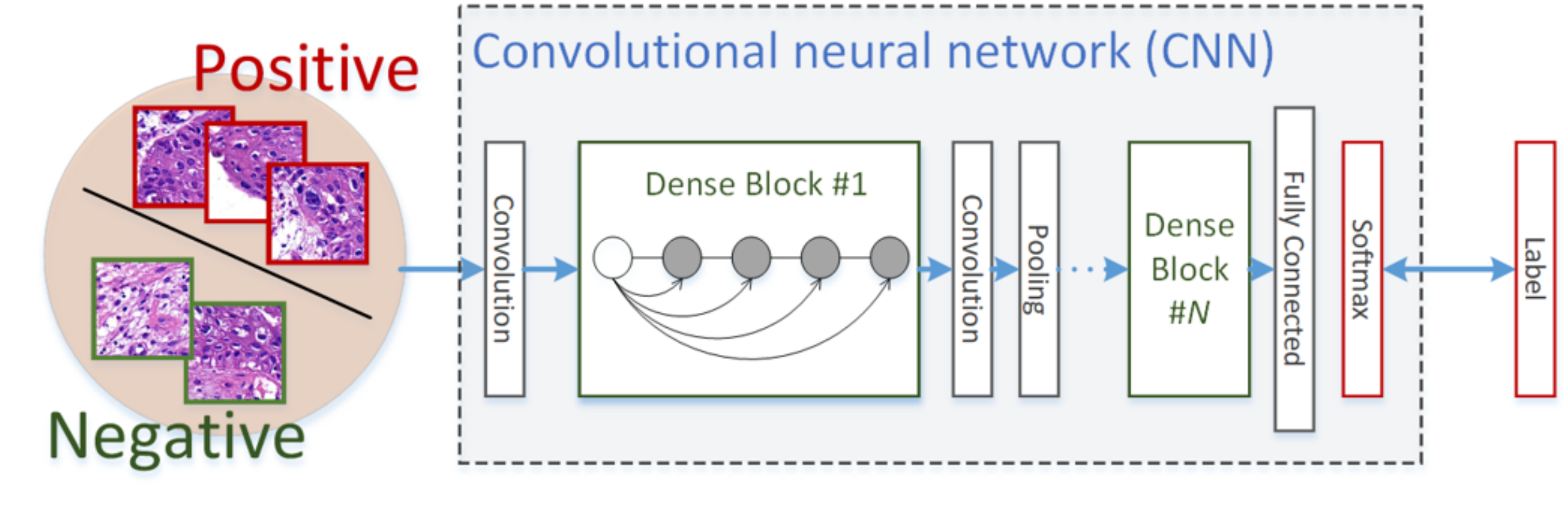}  
}     
\subfigure[pixel-wise segmentation based on FCN.] { 
\label{fig:b}     
\includegraphics[width=12cm]{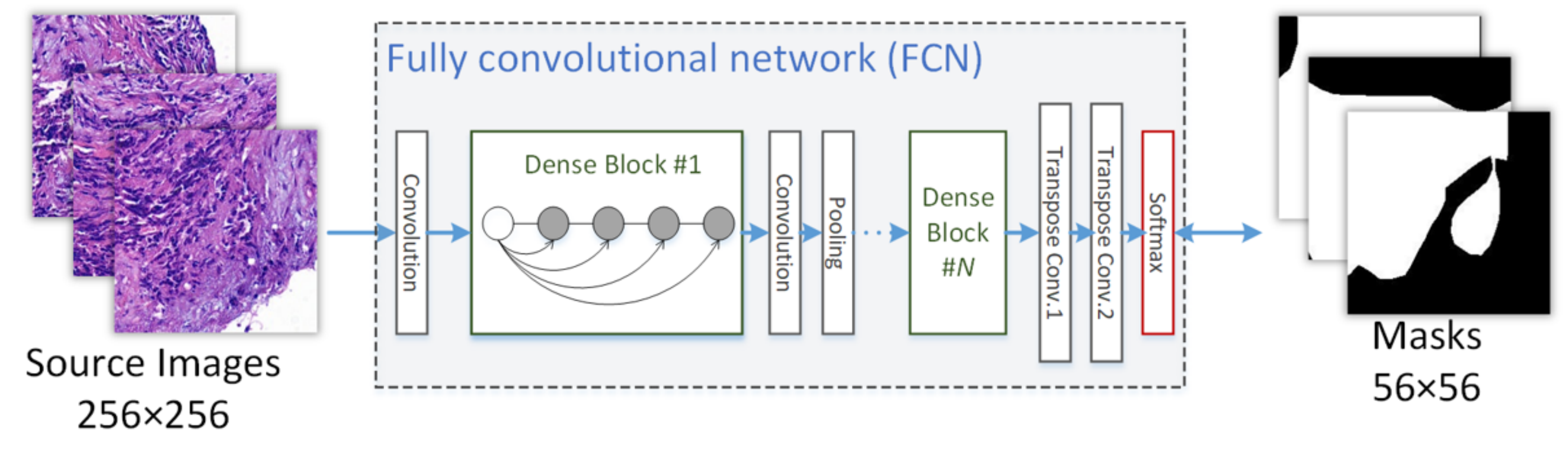}     
}    
\caption{ The network architecture proposed by rank \#4 team.}     
\label{Fig 4}     
\end{figure*}

The rank \#4 team combined advantages of a CNN and a fully convolutional network (FCN) \cite{41} to improve the accuracy of segmentation. At first, a bounding box was manually annotated to locate the tissue regions. CNN was based on DenseNet-121 structure \cite{42} with two output neurons, and the FCN was based on the DenseNet structure consisting of three dense blocks. The first dense block was with five convolutional layers, and the other two were with eight convolutional layers. The architecture of their model was shown as Fig.\ref{Fig 4}. Intel Core i7-7700k CPU and a GPU of Nvidia GTX 1080Ti were used for training. They used cross-entropy with softmax output as the loss function, and Adam as the optimizer for CNN structure. The dice loss and focal loss were set as loss function, and the SGD with momentum was set as the optimizer for FCN structure. 

\begin{figure*}[htpb]
\centerline{\includegraphics[width=15.5cm]{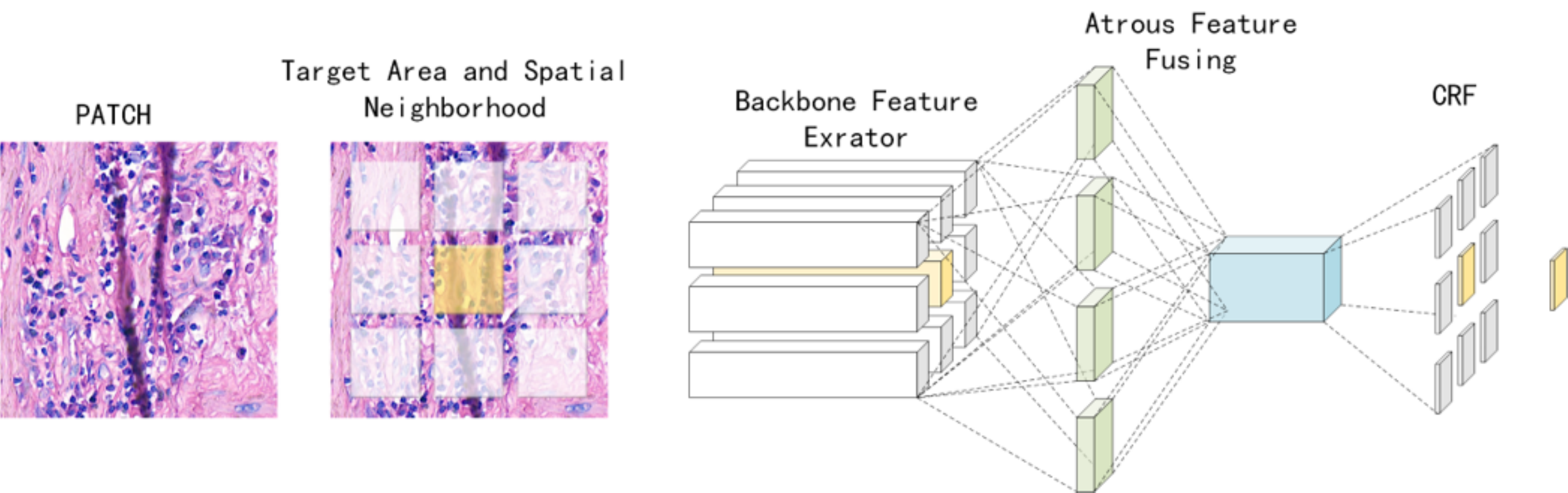}}
\caption{The architecture of model rank \#5 team used.}
\label{Fig 5}
\end{figure*}

The rank \#5 team integrated the Atrous fusing module and CNN feature extractor to build their networks (See Fig.\ref{Fig 5}). They combined ResNet and Inception V2 (IncRes), which replaced eight middle blocks of ResNet18 with Inception’s module. The WSI was split into big patches (with size of $768*768*3$) in the data pre-processing step, and nine small patches were extracted uniformly from each big patch. After feature extraction using IncRes, the multi-atrous convolution was used for feature fusion \cite{43}. The big patches were assigned to TUMOR, NORMAL,and MIX according to the annotation. They mixed MIX patches into TUMOR and NORMAL to keep the balance of the training data. In their experiments, four parallel atrous convolution modules were used to fuse all features with different dilation ratios. The Convolutional Conditional Random Field (CRF) \cite{44} after the concatenate layer was connected. The CRF did not involve in the training stage, but used to modify the output results. They used four NVIDIA GTX 1080Ti 12GB GPU and set the learning rate to 1e-3 for beginning 40 epochs, 7e-4 for the last 20 epochs. The loss function was set as BCEWithLogitsLoss. They illustrated that the model combining IncRes, atrous convolution module, and CRF gave the best segmentation performance. 

The rank \#8 team used a fast deep learning-based model. They put all training sets into the FCN \cite{41} in the AI Explore platform \cite{45}. After training, the test set was tested by the AI Explore platform for real-time lung whole slide segmentation. They used NVIDIA GeForce GTX 1080 Ti to train the model and the SGD with large momentum to avoid the multiple local gradient minimums. The learning rate set to 1e-10.                   

The rank \#9 team used a classification method by labeling large regions instead of distinct pixels. They trained a ResNet18 model with multiple data augmentation methods. An adaptation of threshold was used for cell detection. The training was on a single NVIDIA GeForce GTX 1060, Adam was used as the optimizer with a learning rate set to 1e-4.

The rank \#10 team processed WSIs in large patches with no overlap to capture more context. They evaluated three alternative networks: Small-FCN-16, Small-FCN-32 and Small-FCN-512. For locating the cancer region rather than exact boundaries, they used 4$\times$4 convolutional filters to increase the receptive field at a different level. They also used Imagenet-FCN to train their model. The training was on the NVIDIA Pascal GPU. Adam optimizer, with a decaying learning rate that started with 1e-4, was used to optimize the weights of these networks. The cross-entropy was set as a loss function. They compared different small networks and selected small-FCN-32 with Imagenet-FCN as their final model.

\subsection{The methods based on multi-model}
In general, the single model is not flexible enough to solve complex problems \cite{46}, such as the segmentation of lung cancer regions. Furthermore, training multiple models could significantly improve the generalized performance than only using single model\cite{46}.

\begin{figure*}[htbp]
\centering
\subfigure[]{
\includegraphics[width=13cm]{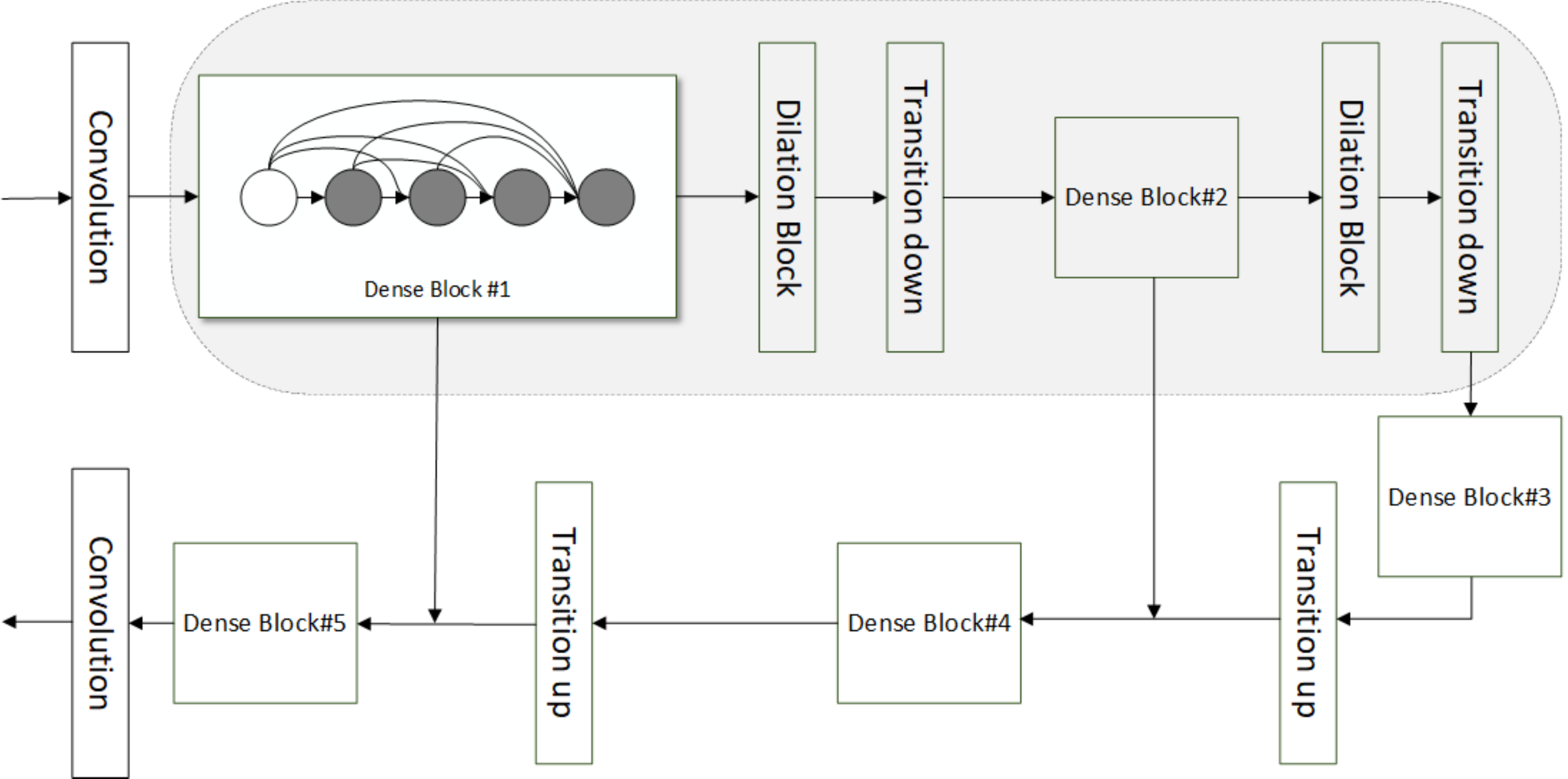}
}
\quad
\subfigure[]{
\includegraphics[width=12cm]{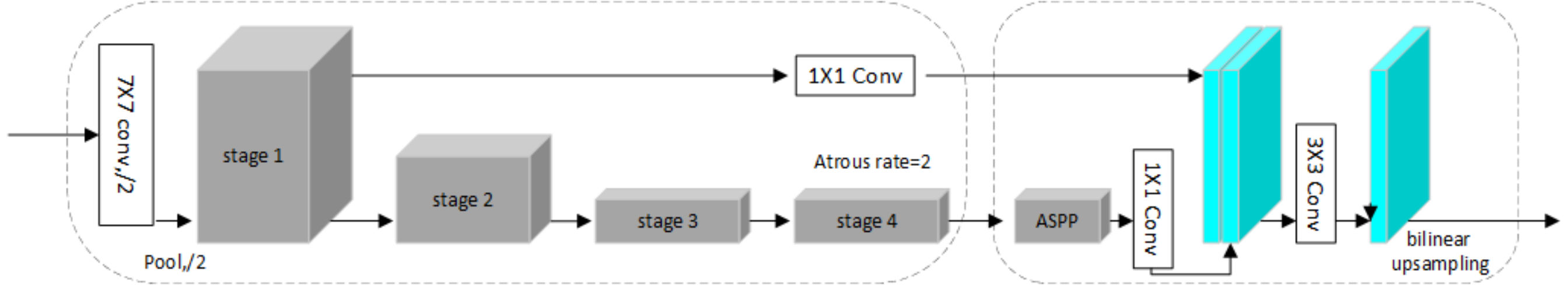}
}
\quad
\subfigure[]{
\includegraphics[width=14cm]{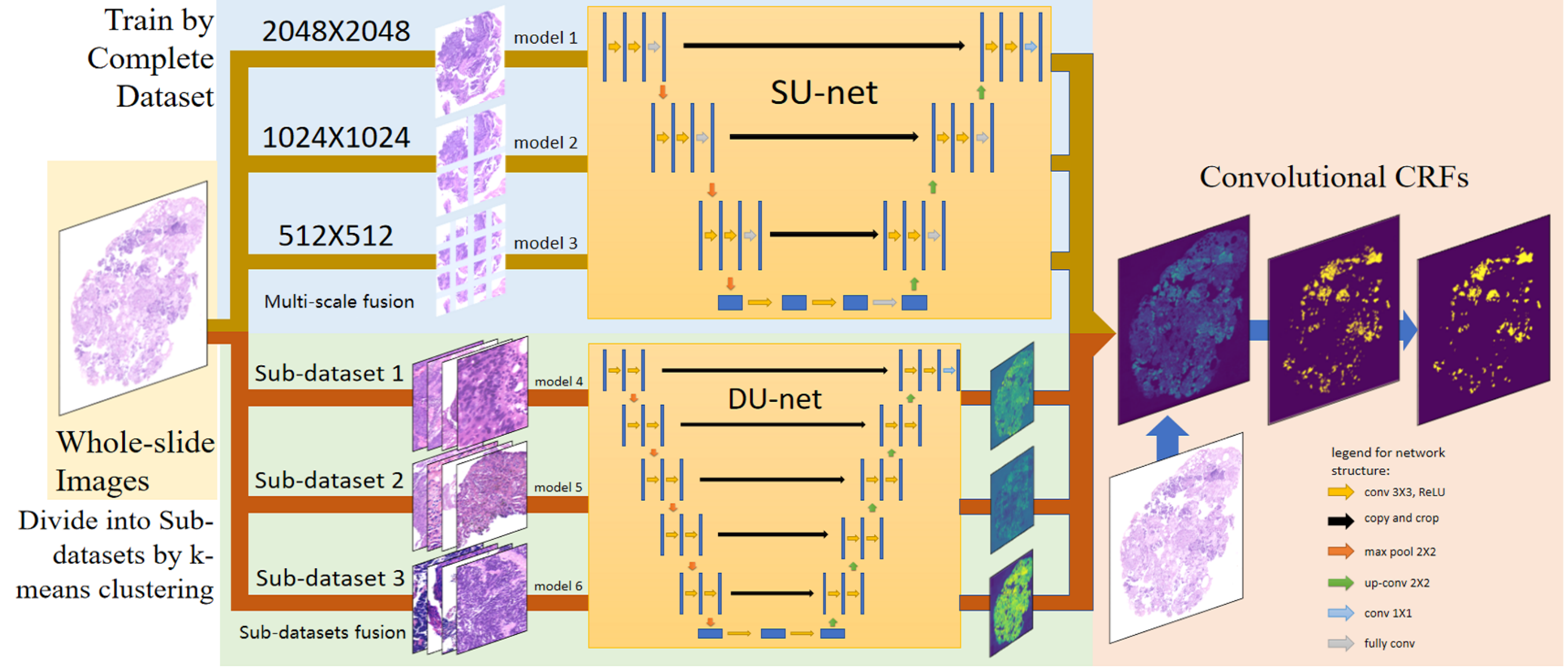}
}
\caption{The network architectures proposed by top three teams.(From top to bottom: rank \#1, rank \#2 and rank \#3)}
\label{Fig 6}
\end{figure*}

The rank \#1 team combined the DenseNets and dilation block to work with U-net. DenseNet \cite{42} connected each layer to every other layer in a feeg-forward fashion (See Fig.\ref{Fig 6}(a)). The U-Net has an encoder-decoder structure with skip connections that enables efficient information flow \cite{47}.  In the dilation block, with the same convolution kernel size, different dilation rates could be utilized to obtain multi-scale features and more context information. The dilation rate (1, 3, 5) with 3x3 kernel were concatenated as the input of the convolution. The dense block was constructed by four layers. They trained different models by changing the loss function through weights and choose the best-performing model to ensemble. This model was more sensitive to tiny lesions and able to capture more context information and multi-scale feature. They used four GPU on Tesla M60 and Adam optimization with default parameters $\beta_1=0.9$, $\beta_2=0.999$ for training, set the initial learning rate to 2$\ast$10-4, and then divided learning rate by 20 in every 20 epochs. The loss function was a combination of dice function and cross-entropy. 


The rank \#2 team refined labels by removing the background within the tumor area and performed data augmentation at the training step. The ResNet50 was used as an encoder network to extract semantic information. The DeepLab V3+ was used for upsampling. They also modified the ResNet architecture to adapt to the task as described below: 1) Down-sampling step in stage4 was eliminated by changing first convolution layer stride 2 to 1; 2) All convolution layers in stage4 had been altered to use atrous rate 1 to 2; 3) Global average pooling layer was removed and attached DeepLab V3+ decoder;4) All convolution layers in DeepLab V3+ decoder used separable convolution. The model is shown as Fig.\ref{Fig 6}(b). In the experiments, they used ImageNet pre-trained weights for encoder and Adam as the optimizer, set the initial learning rate to 1e-4. The loss function was a combination of the cross-entropy loss and soft dice loss. They trained CNN models with five-fold cross-validation and ensembled five models from cross-validation training. 


The rank \#3 team proposed a multi-scale U-net fusion model with the CRF \cite{44} (See Fig.\ref{Fig 6}(c)). The framework fused networks in two ways: multi-scale fusion and sub-datasets fusion. In multi-scale fusion, three models were trained on the whole training set of three resolutions (576, 1152, and 2048 pixels). The network structure was a modified U-net called SU-net (shallow U-net), which focused on the local details of tumor cells. They removed one downsampling and one upsampling steps in the original U-net and added a fully connected layer before every remaining downsampling and upsampling steps. The SU-net included three times of downsampling and upsampling, consisting of 24 layers in total. In sub-datasets fusion: the dataset was divided into three sub-datasets by k-means algorithm \cite{50}. Each sub-dataset was in the same image resolution of 512 pixels and trained on a DU-net (deep U-net model). The DU-net added one additional downsampling and upsampling stages, consisting of 28 layers. In the experiments, they used the soft-max combining with the cross entropy loss as the loss function. 

\begin{figure*}[htpb]
\centerline{\includegraphics[width=8.5cm]{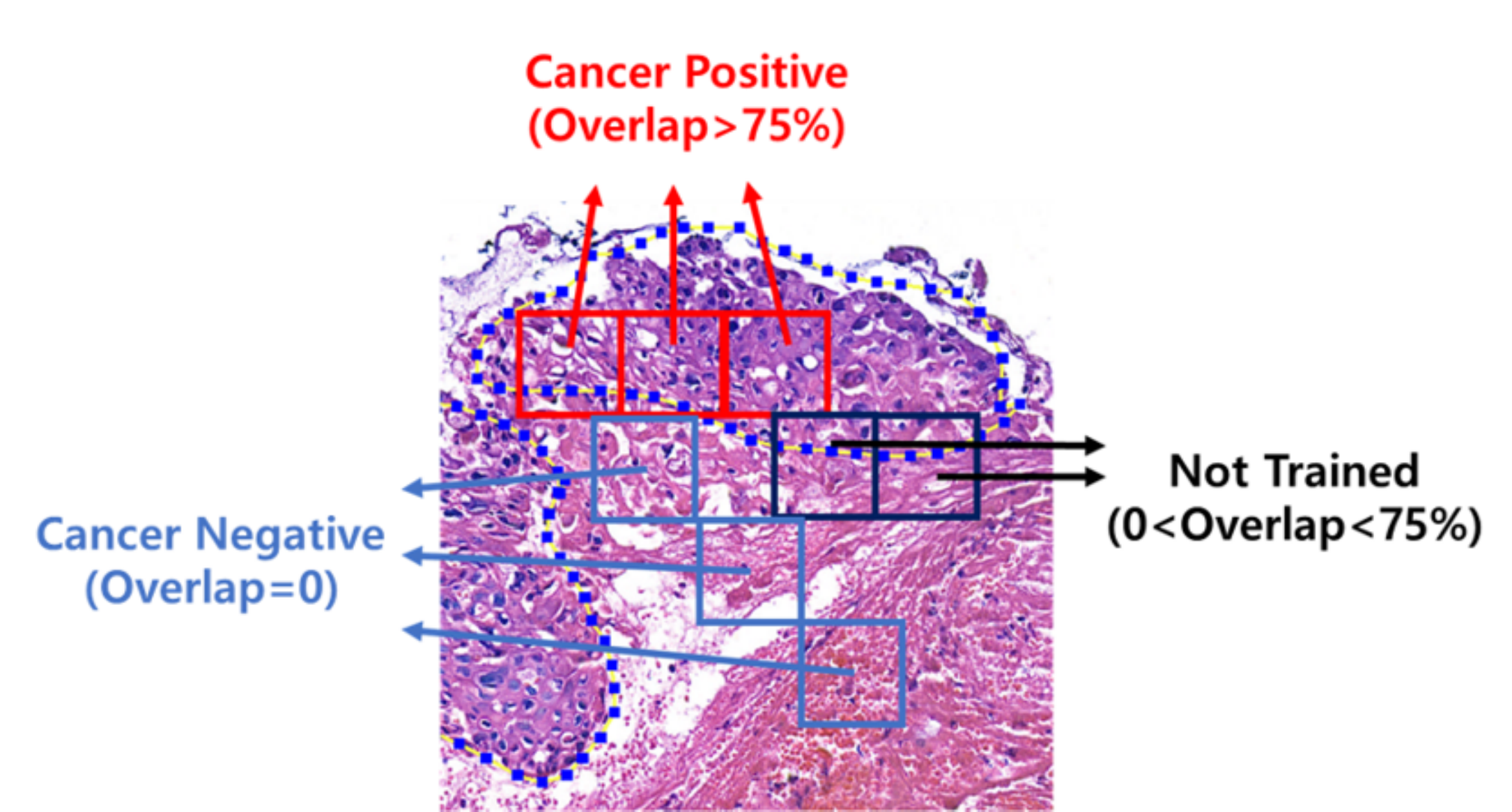}}
\caption{ Tile labeling strategy of rank \#6 team.}
\label{Fig 7}
\end{figure*}

The rank \#6 team used of existing classical models, including ResNet101, ResNet152, DenseNet201, DenseNet264 and Mdrn80 (a short version of the network that DeepMind \cite{48}). They used the tile labeling strategy to label the cancer regions (See Fig.\ref{Fig 7}). Tile overlapped more than 75\% with annotated cancer region was defined as a positive cancer tile, and the tile without overlapping the cancer region was a negative tile. Other tiles were not used for training. They trained and ensemble 16 models to conduct the experiments. They used three NVIDIA RTX Titan GPUs and Adam optimization with a learning rate of 1e-4 for training. Cross-entropy was set as the loss function.
 
\begin{figure*}[ht]
\centerline{\includegraphics[width=10cm]{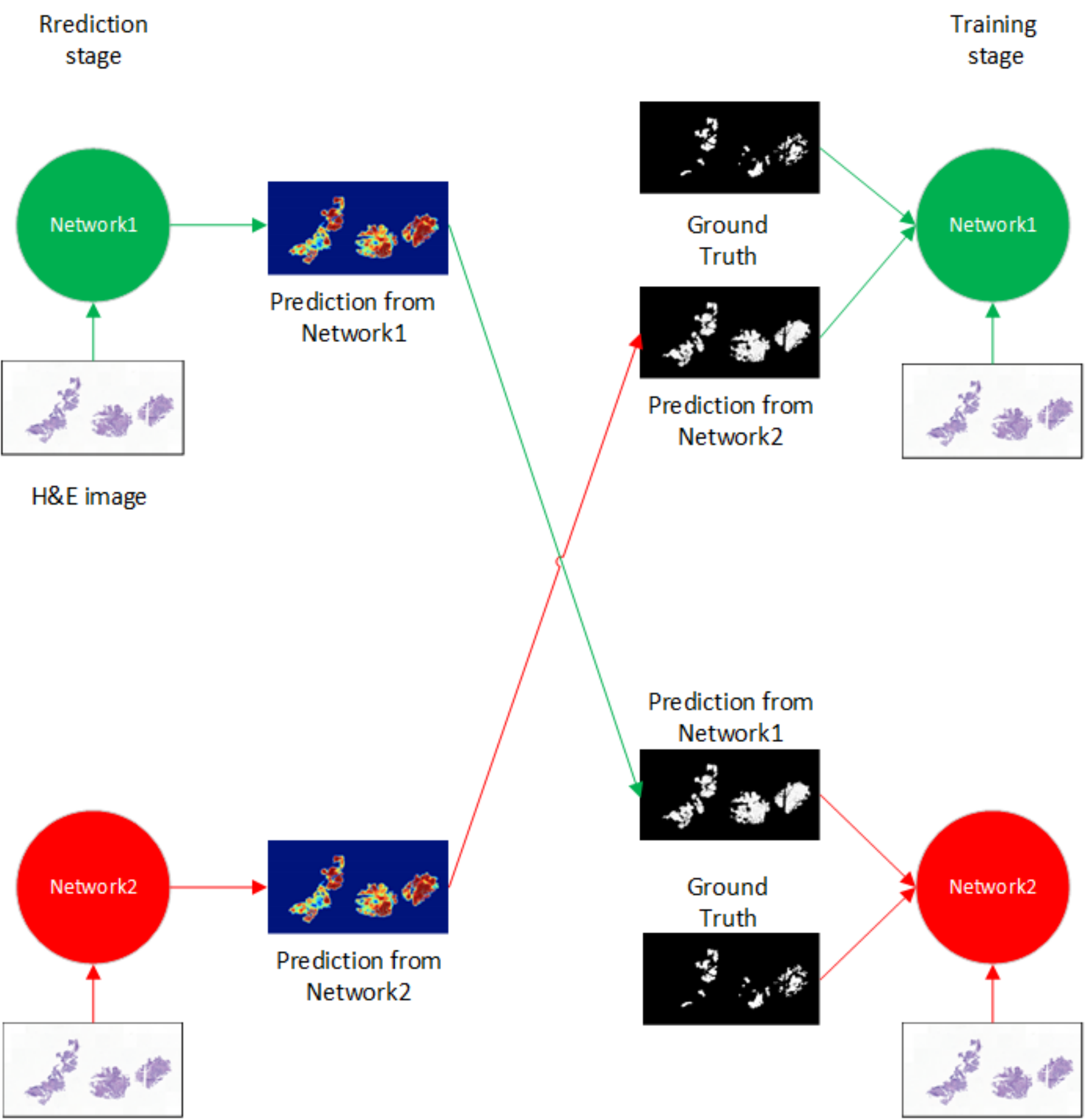}}
\caption{Schematic diagram of pixel-level Co-teaching algorithm.}
\label{Fig 8}
\end{figure*}

The rank \#7 team used “Co-teaching” to train networks (See Fig.\ref{Fig 8}). Co-teaching \cite{49} aims to clean the noisy label. The proposed method trained two networks simultaneously. In each mini-batch of data, each network viewed its small-loss instances as useful knowledge and taught such instances to its peer network for updating the parameters. Comparing with the original Co-teaching algorithm, the main difference was the dynamic drop rate $\Re(T)$, which controlled the number of clean-instances selected for training. It was used to avoid the training error from a network to be directly transferred back on itself (See Algorithm.\ref{alg: Pixel-level Co-teaching Algorithm}). They used the fully convolutional DenseNet (FC-DenseNet) 103 network as a backbone. The two FC-DenseNet 103 networks were trained from scratch simultaneously using the same data. In their experiment, the networks were trained using four NVIDIA GeForce GTX 1080 Ti GPUs, and an Adam optimizer was used with an initial learning rate of 1.5e-4.
\begin{algorithm}[ht]
\renewcommand{\algorithmicrequire}{ \textbf{Input:}} 
\renewcommand{\algorithmicensure}{ \textbf{Output:}} 
\caption{ Pixel-level Co-teaching Algorithm}
\label{alg: Pixel-level Co-teaching Algorithm}
\begin{algorithmic}[1]
\REQUIRE $w_{f}$ and $w_{g}$,learning rate $\eta$,epoch $T_{max}$,iteration $N_{max}$;
\FOR{T=1,2,...,$T_{max}$}
\STATE Shuffle the training set D; 
\FOR{N=1,2,...,$N_{max}$}
\STATE Fetch J and L from D;
\STATE Obtain $\overline{L}_{f}=\sigma[f_{w_{f}}(J)]>0.5$;
\STATE Obtain $\overline{L}_{g}=\sigma[f_{w_{g}}(J)]>0.5$; 
\STATE Update $w_{f}=w_{f}-\eta\nabla l_{f}(f,\overline{L}_{g},L)$
\STATE Update $w_{g}=w_{g}-\eta\nabla l_{g}(f,\overline{L}_{f},L)$ 
\ENDFOR
\ENDFOR
\ENSURE $w_{f}$ and $w_{g}$
\end{algorithmic}
\end{algorithm}

A detailed description of the top 10 methods will be uploaded to our challenge website\footnote{\url{http://acdc-lunghp.grand-challenge.org}}. 

\section{Result and Discussion}

\subsection{Comparisons of Top 10 Methods}

\begin{figure*}[ht]
\centerline{\includegraphics[width=18cm]{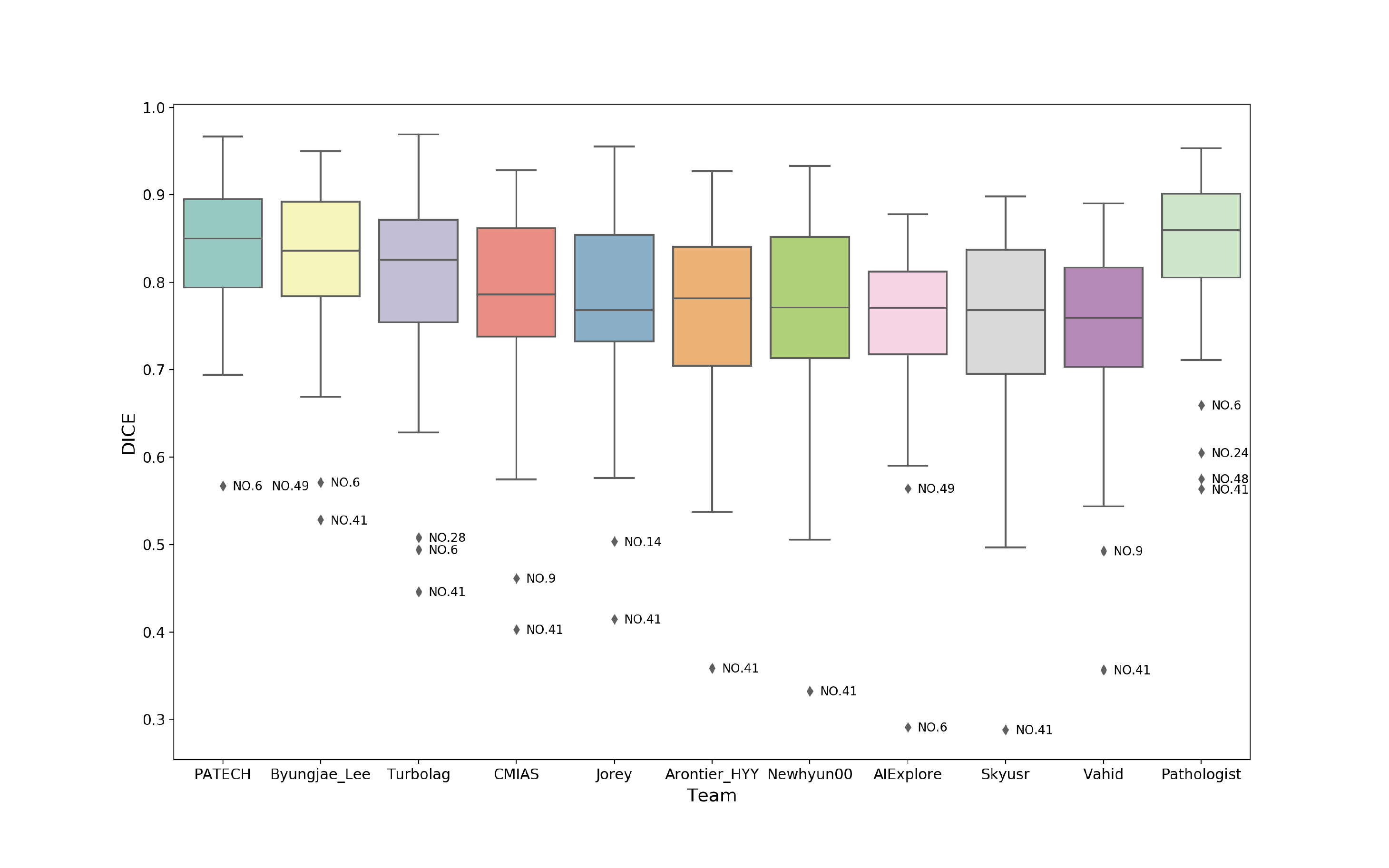}}
\caption{Comparisons of top 10 teams on the test set.}
\label{11}
\end{figure*}

\begin{figure*}[ht]
\centering
\subfigure[]{
\includegraphics[scale=0.55]{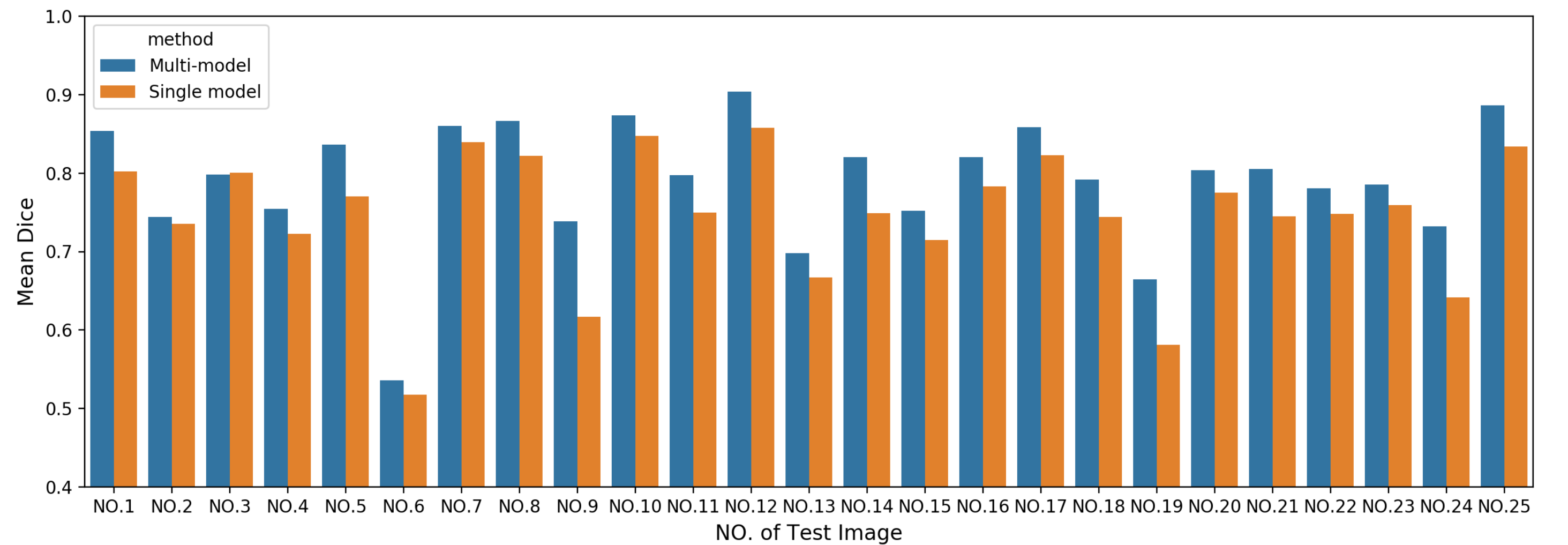}
}
\quad
\subfigure[]{
\includegraphics[scale=0.55]{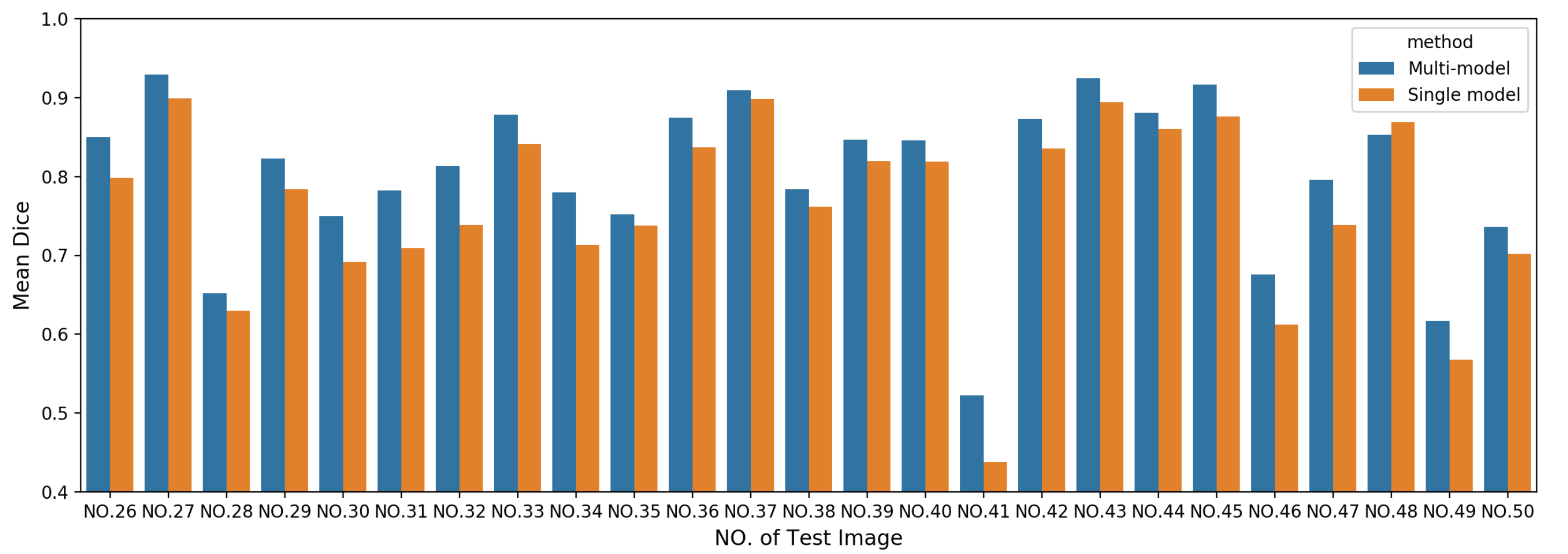}
}
\caption{The mean DC of multi-model methods and single model methods for all 50 test image. (a) test image NO.1-NO.25.(b) test image NO.26-NO.50}
\label{12}
\end{figure*}

The box-plot of the DC for test set of the top 10 teams was shown in Fig.\ref{11}. The inter-observer variability between the two pathologists was also assessed using the mean DC, which was 0.8398 (See Fig.\ref{11}). And mean DC of multi-model methods and single model methods for each test image was shown in Fig.\ref{12}. 

\begin{figure*}[ht]
\centering
\subfigure[]{
\includegraphics[width=6cm,height=6cm]{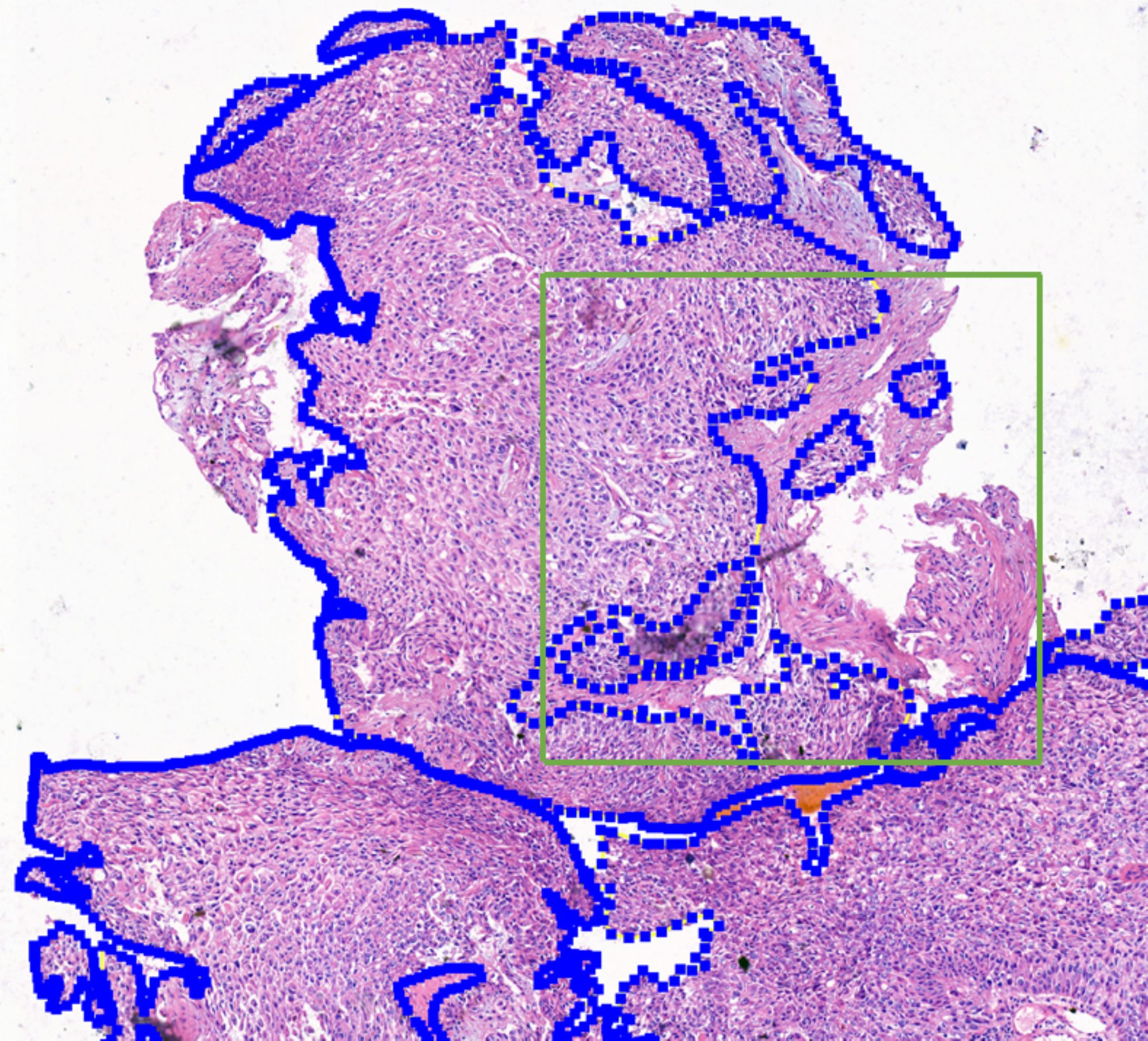}
}
\quad
\subfigure[]{
\includegraphics[width=6cm,height=6cm]{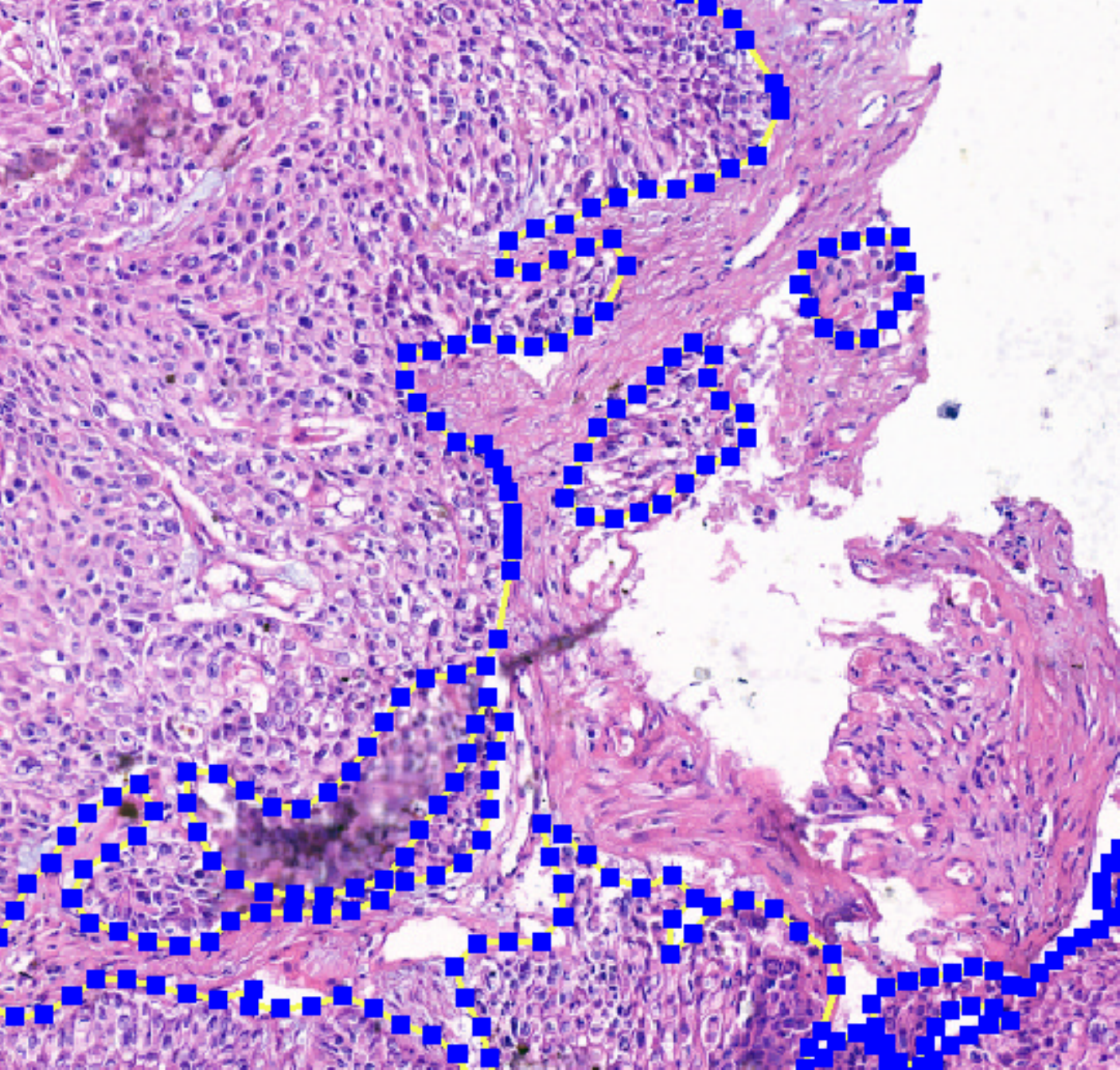}
}
\quad
\subfigure[]{
\includegraphics[width=6cm,height=6cm]{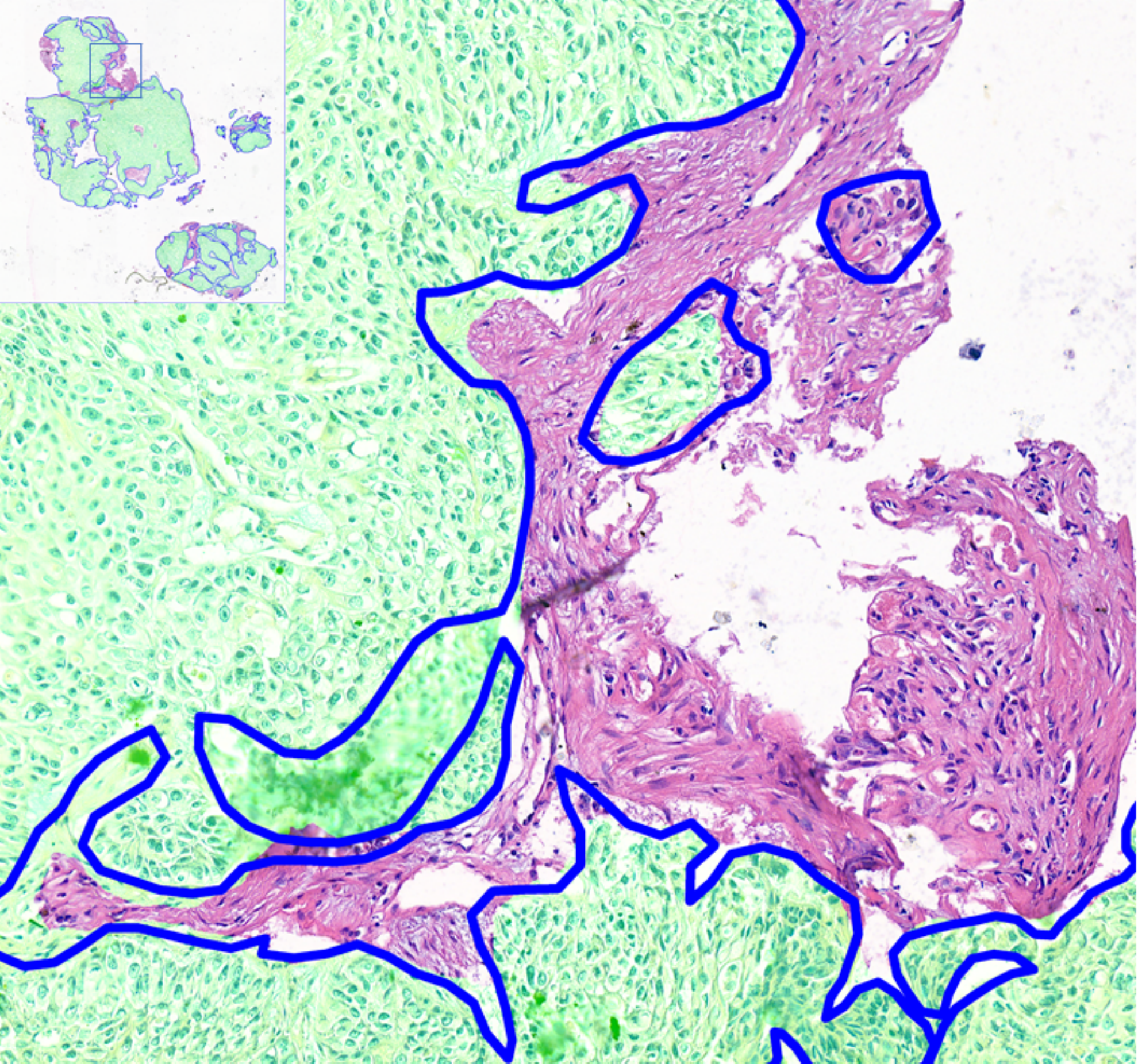}
}
\quad
\subfigure[]{
\includegraphics[width=6cm,height=6cm]{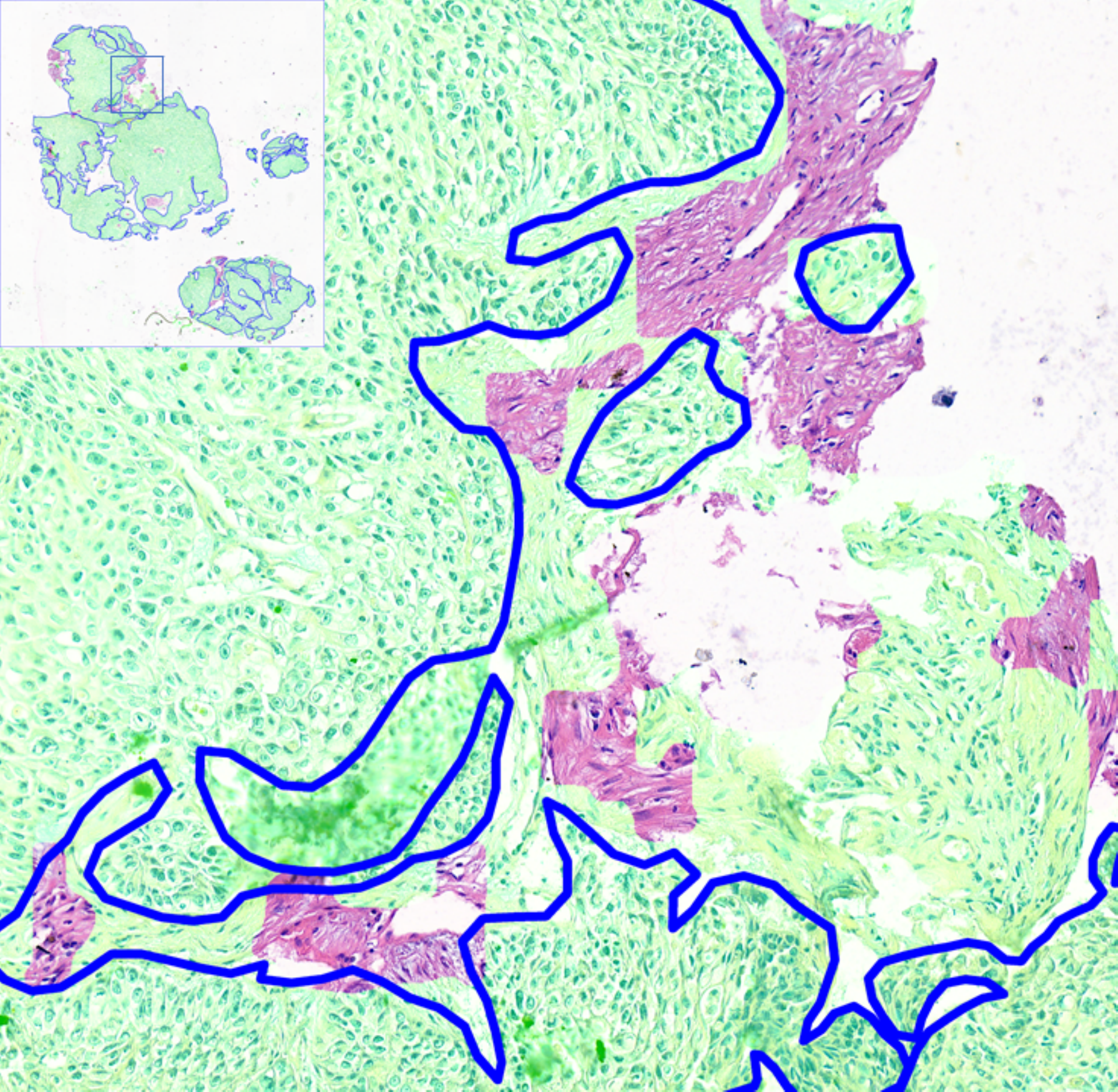}
}
\caption{ Pathological WSI for test image NO.27. (a) image with annotation (blue line). (b) selected patch of (a). (c) and (d) results of rank \#2 team (DC = 0.9435) and rank \#8 team (DC = 0.8653).}
\label{13}
\end{figure*}

All teams got a relative high DC on the NO.27 test image. The DC ranged from 0.8653 to 0.9435. This sample was well prepared during H\&E staining like most of the training datasets, and the cancer tissue was clearly shown in this image. One could see typical results from two different teams in Fig.\ref{13}(c) and (d). The tissue was shown in Fig.\ref{13}(a) and (b). 

\begin{figure*}[ht]
\centering
\subfigure[]{
\includegraphics[width=6cm,height=6cm]{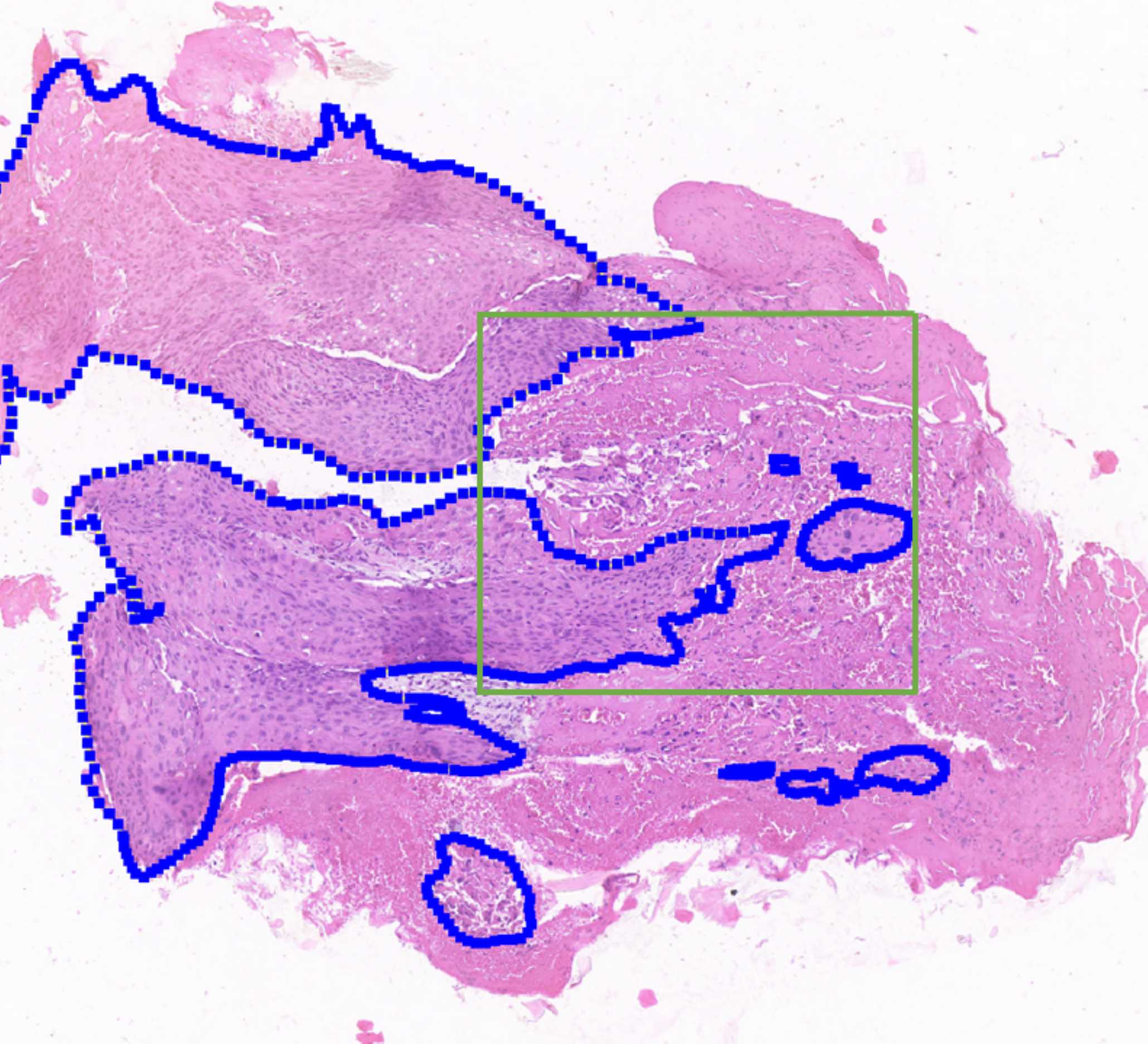}
}
\quad
\subfigure[]{
\includegraphics[width=6cm,height=6cm]{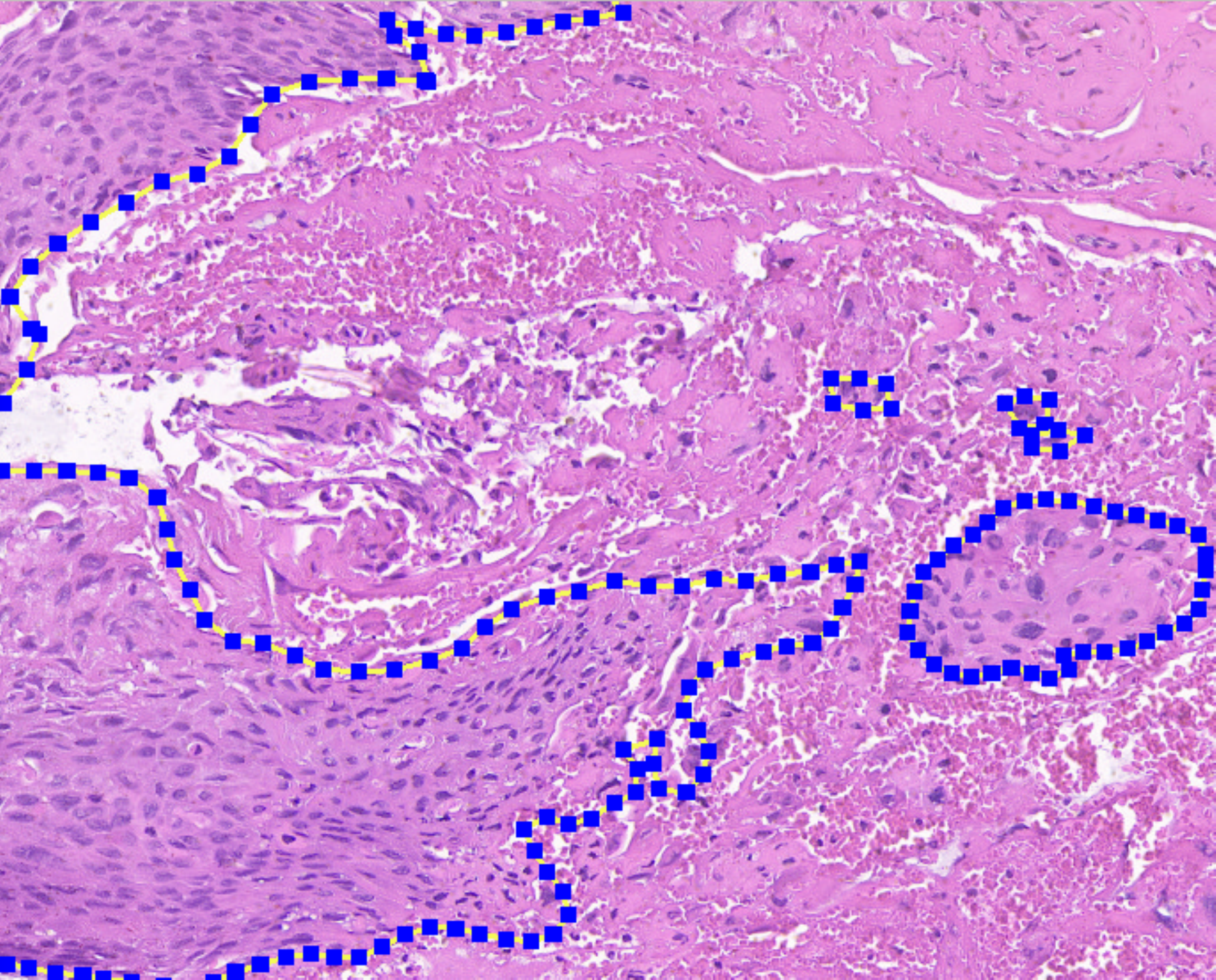}
}
\quad
\subfigure[]{
\includegraphics[width=6cm,height=6cm]{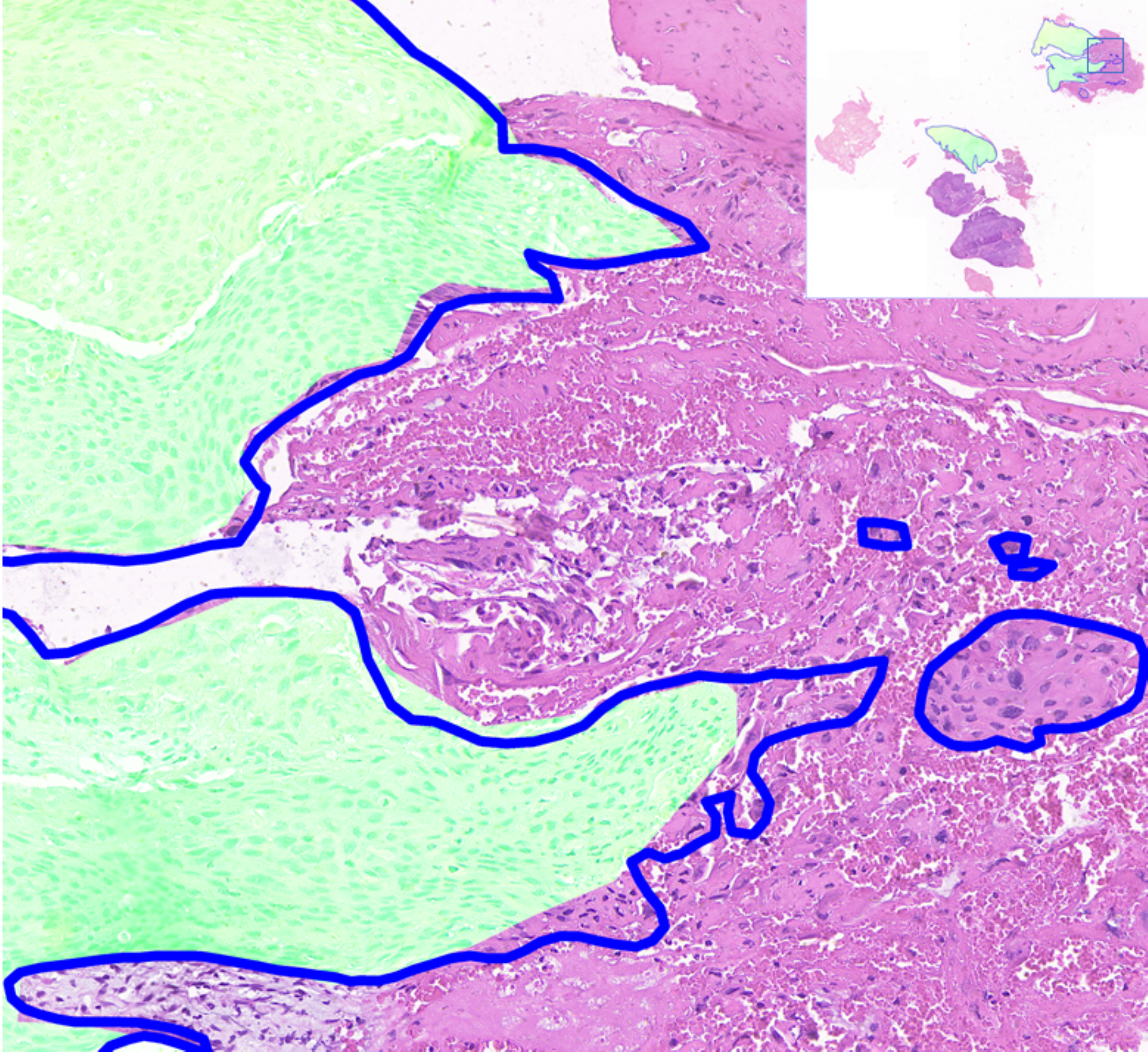}
}
\quad
\subfigure[]{
\includegraphics[width=6cm,height=6cm]{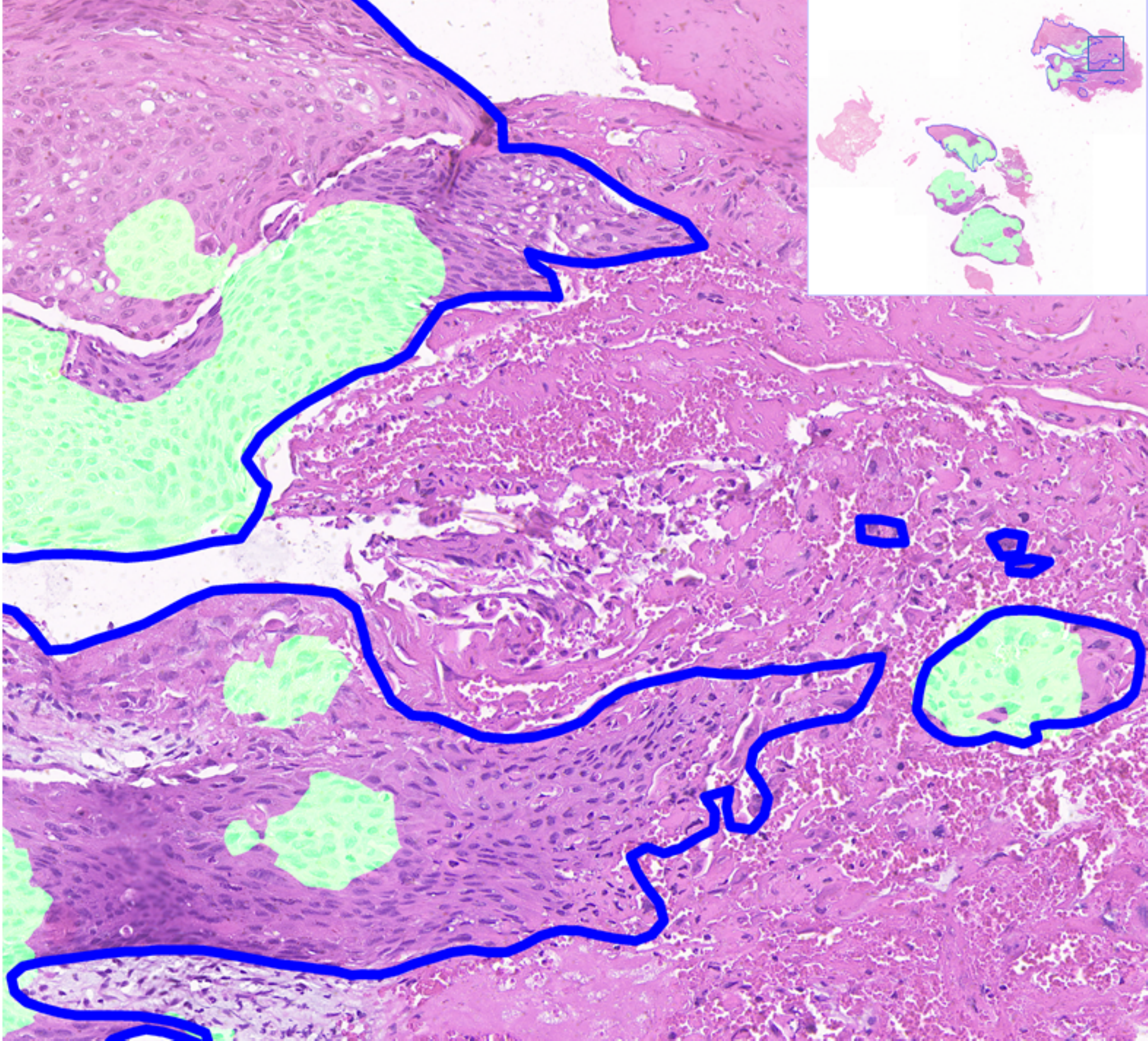}
}
\caption{ Pathological WSI for test image NO.41.  (a) image with annotation (blue line). (b) selected patch of (a). (c) and (d) results of rank \#1 team (DC = 0.9458) and rank \#7 team (DC = 0.3327). }
\label{14}
\end{figure*}

In contrast, most of the teams got relative low DC on the NO.41 test image (between 0.2-0.5, see Fig.\ref{11}). Only rank \#1 team got a high DC (0.9458), among others. A visual comparison could be found in Fig.\ref{14}(c) and (d). NO.41 was an example of high differentiated squamous cell carcinoma. The abnormal cells were similar to normal cells in this case. And the color appearance of this slide was not consistent with other slides due to the off-standard H\&E staining process. Models from most of the teams might not be generalized enough to deal with this problem.

In order to better evaluate the performance of top 10 methods on the test set, we listed the mean DC on the images of the squamous cell carcinoma (SCC), small cell carcinoma (SCLC), and adenocarcinoma (ADC) (See Table \ref{tab10}). The result illustrated that the accuracy of the segmentation depends on how the cancer cells grow. There were no other components in the squamous cell carcinoma nest, so the segmentation accuracy was higher than the other two types. However, small cell carcinoma spread along the sparse fibrous interstitium and gaps, and its cytoplasm was minimal. The adhesion between the cells was inferior, and it was easy to loosen, plant, transfer. Also, the cells were squeezed and deformed during the biopsy, resulting in unclear boundaries. So high performance was hard to be achieved for SCLC. ADC grew along the alveolar wall, and there were too many vascular interstitial components that may affect the segmentation accuracy.

\begin{table}[]
\centering
\caption{The comparisons of multi-model and single model methods on three types OF Lung cancer.}\label{tab10}
\begin{tabular}{|c|c|c|c|}
\hline
                      & \textbf{SCC} & \textbf{SCLC} & \textbf{ADC} \\ \hline
\textbf{Multi Model}  & 0.8205       & 0.7521        & 0.7888       \\ \hline
\textbf{Single Model} & 0.7797       & 0.7186        & 0.7468       \\ \hline
\textbf{All}          & 0.8001       & 0.7353        & 0.7678       \\ \hline
\end{tabular}
\end{table}

\subsection{Multi-model V.S. Single model}
The sign rank test was used to evaluate differences of DC between multi-model and single model methods (based on Fig.\ref{12}). The multi-model methods gave significantly better results (\textit{p}=1.0872e-09) than single model methods. Besides comparing the DC, we also calculated accuracy, FPR, and FNR of detection for the top 10 methods (see Table \ref{tab7}).

We can see from Table \ref{tab7} that FNR of multi-model methods were generally lower than single model methods. We can see that different types of cancer tissue were with different appearances. The current challenge may difficult to provide enough data for all types of cancer. Using a single model might not be sufficient in identifying specific types of cancer. Through model fusion, we could combine multiple models' performance and reduce the probability of missed inspections.

\begin{table*}[ht]
\centering
\begin{threeparttable}[b]
\caption{Quantitative comparisons of multi-model and single model methods on test set.}\label{tab7}
\begin{tabular}{|c|c|c|c|c|c|}
\hline
                                                                                 & \textbf{Rank} & \textbf{Mean.DC}     & \textbf{Accuracy} & \textbf{\begin{tabular}[c]{@{}c@{}}FNR\\ (False Negative Rate)\end{tabular}} & \textbf{\begin{tabular}[c]{@{}c@{}}FPR\\ (False Positive Rate)\end{tabular}} \\ \hline
\multirow{5}{*}{\textbf{\begin{tabular}[c]{@{}c@{}}Multi \\ Model\end{tabular}}} & 1             & \textbf{0.8372$\pm$0.0858} & \textbf{0.9505}   & \textbf{0.0948}                                                              & 0.0469                                                                       \\ \cline{2-6} 
                                                                                 & 2             & \textbf{0.8297$\pm$0.0867} & \textbf{0.9508}   & \textbf{0.1372}                                                              & \textbf{0.0391}                                                              \\ \cline{2-6} 
                                                                                 & 3             & \textbf{0.7968$\pm$0.1081} & \textbf{0.9462}   & 0.1531                                                                       & \textbf{0.0415}                                                              \\ \cline{2-6} 
                                                                                 & 6             & 0.7638$\pm$0.1107          & 0.9289            & 0.1442                                                                       & 0.0596                                                                       \\ \cline{2-6} 
                                                                                 & 7             & 0.7552$\pm$0.1237          & 0.9307            & 0.1801                                                                       & 0.0511                                                                       \\ \hline
\multirow{5}{*}{\textbf{\begin{tabular}[c]{@{}c@{}}Single\\ Model\end{tabular}}} & 4             & 0.7700$\pm$0.1177          & 0.9375            & 0.1997                                                                       & \textbf{0.0433}                                                              \\ \cline{2-6} 
                                                                                 & 5             & 0.7659$\pm$0.1130          & 0.9369            & 0.1849                                                                       & 0.0486                                                                       \\ \cline{2-6} 
                                                                                 & 8             & 0.7510$\pm$0.0973          & 0.9231            & \textbf{0.1404}                                                              & 0.0721                                                                       \\ \cline{2-6} 
                                                                                 & 9             & 0.7465$\pm$0.1188          & 0.9319            & 0.2328                                                                       & 0.0437                                                                       \\ \cline{2-6} 
                                                                                 & 10            & 0.7354$\pm$0.1149          & 0.9213            & 0.1538                                                                       & 0.0684                                                                       \\ \hline
\end{tabular}
\begin{tablenotes}
\centering
     \item[1] The top 3 methods were shown in bold format.
   \end{tablenotes}
  \end{threeparttable}
\end{table*}
\subsection{Pre-trained model V.S. No pre-trained model }
Transfer learning is a commonly used method in the AI community. Using the pre-trained model for fine-tuning can reduce training time and achieve better results in several applications. Three teams used ImageNet pre-trained weights to initialize their models. In the challenge, the methods using pre-trained models did not outperform the method that learning from scratch. This might be because the digital pathology domain is inherently different from the ImageNet domain.

The CAMELYON16, TUPAC, and CAMELYON17 challenges aimed at detecting the micro- and macro- metastases in the lymph node in H\&E stained WSIs (CAMELYON16/17) and assessing tumor proliferation in breast cancer (TUPAC). Using these data for pre-training might get good results. However, none of the teams used a pre-training model from those data. 
\subsection{Label Refine}
\begin{figure}[ht]
\centerline{\includegraphics[width=8cm,height=8cm]{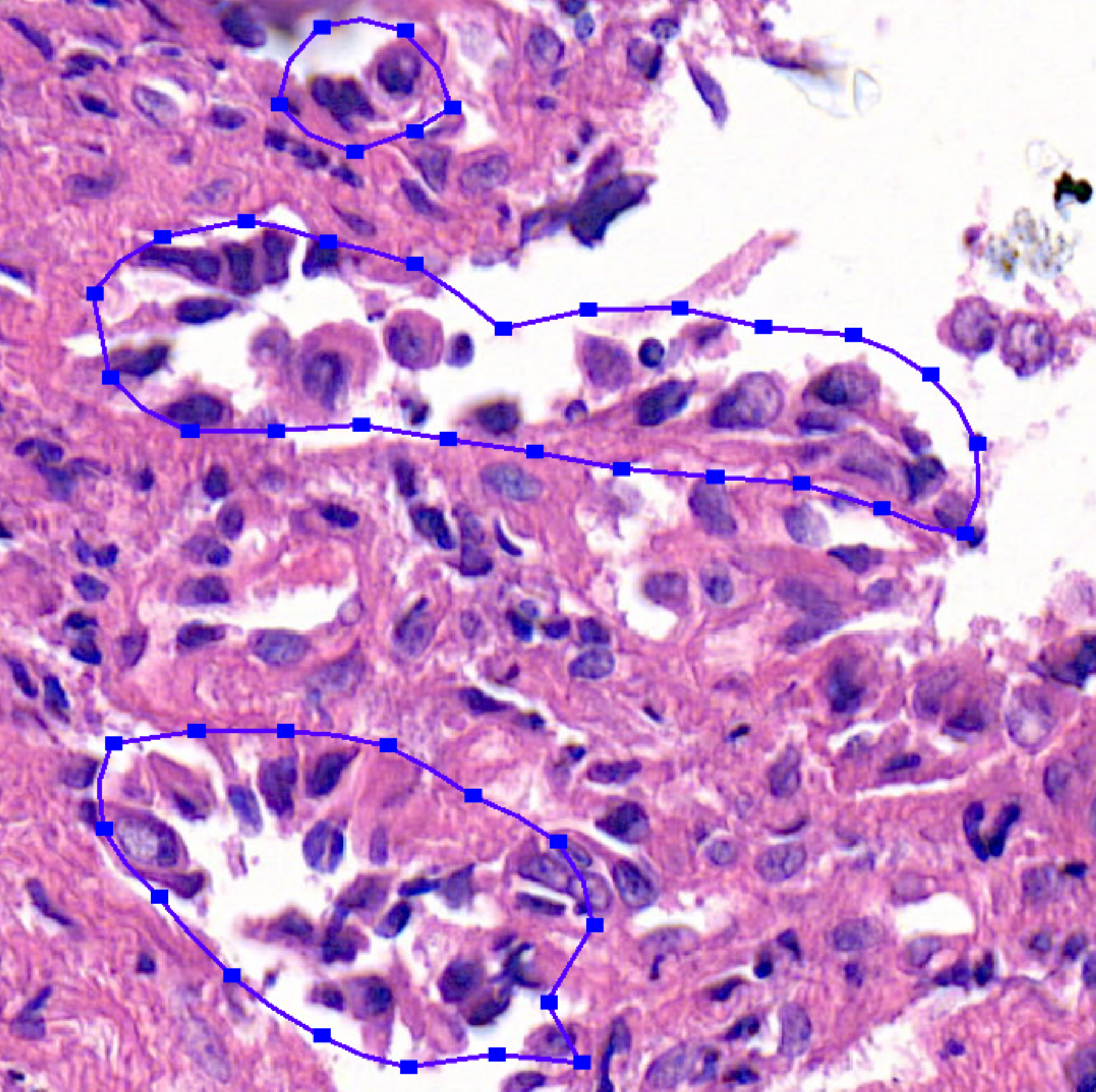}}
\caption{ The sample with label noise (annotation contains background region).}
\label{15}
\end{figure}

Experienced pathologist annotated cancer regions using ASAP software. We intended to make relative rough labels for the training set (e.g., label contains background region shown as Fig.\ref{15}) to evaluate the robustness of the methods in dealing with label noise. All background and not labeled tissue were kept in the training set as well. It makes tumor tissue and the normal area extremely unbalanced in the training set. Therefore, label refine is one of the significant issues that should be taken into consideration in this challenge to keep data balanced.

Several teams used different methods to refine the label. Three teams (rank \#1, \#2, \#5) used the Otsu algorithm to remove the background area in tumor tissue labeled and obtain a tighter boundary of the cancer region. The team (rank \#4) located the tissue region by a bounding box and filtered the blank areas using a threshold. The team (rank \#6) used a tile labeling strategy in their method and removed background by the percentage of pixel values above 200 in grayscale space. The team (rank \#7) used the "Co-teaching" algorithm to refine noisy annotation. The team (rank \#10) increased the receptive field at a different level, and they tried to label regions of WSIs rather than finding the exact boundaries.

We found that teams using the Otsu algorithm that removing background gave relatively higher DC. The preprocess for removing the label noise (such as the background in the label area) is essential for model training for the challenge despite the network design. 

\section{CONCLUSION}
In this paper, the ACDC@LungHP challenge was summarized. The current stage of the challenge focused on lung cancer segmentation. 200 slides were used for this challenge, and methods from the top 10 teams were selected for comparison. In general, multi-model method was relatively better than single model-based methods. The results showed the potentiality
of using deep learning for accurate lung cancer diagnosis on WSI. 

All submitted methods were based on deep learning, but the networks were quite different. Methods based on multi-model outperformed single model method (mean DC of a single model is 0.7544$\pm$0.0991 and multi-model is 0.7966$\pm$0.0898). Unlike fine-tuning for other computer vision tasks, the submitted methods did not benefit too much from the ImageNet pre-trained models. The pre-processing for the label noise during the training stage is crucial since our training data was not accurately labeled for test set.

In the coming second stage of this challenge, we will focus on classifying the primary lung cancer subtypes (e.g., squamous carcinoma, adenocarcinoma) using WSI biopsy. At least 4000 slides collected from multiple medical centers will be released in mid-2020. We believe that the experiences of the first stage will greatly help digital pathology communities to achieve better performance for the second stage.

\section*{acknowledgment}
The authors would like to thank Ping Liu, Jun Tang, Tao Xu, Jun XU, Shanshan Wan, Ke Lou, Hui Li, Keyu Li and Yusheng Yan for collecting all the images.


\begin{thebibliography}{00}

\bibitem{1} Jemal, A., et al.,"Annual Report to the Nation on the Status of Cancer, 1975-2014, Featuring Survival.”J Natl Cancer Inst,vol. 109,no. 9,2017.

\bibitem{2} Kim, H.J., et al.,"Outcome of incidentally detected airway nodules."Eur Respir J.,vol. 47,no. 5,pp. 1510-7,2016. 

\bibitem{3} Andolfi, M., et al.,"The role of bronchoscopy in the diagnosis of early lung cancer: a review."J Thorac Dis,vol. 8,no. 11,pp. 3329-3337,2016.

\bibitem{4} Thomas, J.S., et al.,"How reliable is the diagnosis of lung cancer using small biopsy specimens? Report of a UKCCCR Lung Cancer Working Party." Thorax,vol. 48,no. 11,pp. 1135-9,1993.

\bibitem{5}
Srinidhi, Chetan L, O. Ciga, and A. L. Martel."Deep neural network models for computational histopathology: A survey."arXiv:1912.12378, 2019.

\bibitem{6}
Veta M, Heng YJ, Stathonikos N, et al. Predicting breast tumor proliferation from whole-slide images: The TUPAC16 challenge [published correction appears in Med Image Anal. 2019 Aug;56:43]. Med Image Anal. 2019;54:111‐121. doi:10.1016/j.media.2019.02.012

\bibitem{7}
Zhang, Z., Chen, P., McGough, M. et al. "Pathologist-level interpretable whole-slide cancer diagnosis with deep learning." Nat Mach Intell, vol.1, pp. 236–245, 2019. https://doi.org/10.1038/s42256-019-0052-1

\bibitem{8}
G. Xu et al., "CAMEL: A Weakly Supervised Learning Framework for Histopathology Image Segmentation," 2019 IEEE/CVF International Conference on Computer Vision (ICCV), Seoul, Korea (South), 2019, pp. 10681-10690, doi: 10.1109/ICCV.2019.01078.

\bibitem{9}
Bera, K., Schalper, K.A., Rimm, D.L., Velcheti, V., Madabhushi, A. "Artificial intelligence in digital pathologynew tools for diganosis and precision oncology." Nature Reviews Clinical Oncology, vol.16, pp.703-715, 2019.

\bibitem{10} Litjens, G., et al.,"A survey on deep learning in medical image analysis."Med Image Anal,vol. 42,pp. 60-88,2017. 

\bibitem{11} Shen, D., G. Wu, and H.I. Suk,"Deep Learning in Medical Image Analysis."Annu Rev Biomed Eng,vol. 19,pp. 221-248,2017. 

\bibitem{12} Wang, H., et al.,"Mitosis detection in breast cancer pathology images by combining handcrafted and convolutional neural network features."J Med Imaging (Bellingham),vol. 1,no. 3,p. 034003,2014. 

\bibitem{13} Shkolyar, A., et al.,"Automatic detection of cell divisions (mitosis) in live-imaging microscopy images using Convolutional Neural Networks." Conf Proc IEEE Eng Med Biol Soc, 2015,pp. 743-6.

\bibitem{14} Malon, C.D. and E. Cosatto,"Classification of mitotic figures with convolutional neural networks and seeded blob features."J Pathol Inform, vol. 4,pp. 9,2013.

\bibitem{15} Xie, Y., et al.,"Beyond Classification: Structured Regression for Robust Cell Detection Using Convolutional Neural Network."Med Image Comput Comput Assist Interv,vol. 9351,pp. 358-365,2015.

\bibitem{16} Xie, Y., et al.,"Deep Voting: A Robust Approach Toward Nucleus Localization in Microscopy Images."Med Image Comput Comput Assist Interv,vol. 9351,pp. 374-382,2015.

\bibitem{17}
Xing F, Cornish TC, Bennett T, Ghosh D, Yang L. Pixel-to-Pixel Learning With Weak Supervision for Single-Stage Nucleus Recognition in Ki67 Images. IEEE Trans Biomed Eng. 2019;66(11):3088‐3097. doi:10.1109/TBME.2019.2900378

\bibitem{18} Gao, Z., et al.,"HEp-2 Cell Image Classification With Deep Convolutional Neural Networks."IEEE J Biomed Health Inform,vol. 21,no. 2,pp. 416-428,2017. 

\bibitem{19} Bauer, S., et al. (2016)" Multi-Organ Cancer Classification and Survival Analysis." ArXiv e-prints 1606.

\bibitem{20} H. Chen, X. Qi, L. Yu and P. Heng, "DCAN: Deep Contour-Aware Networks for Accurate Gland Segmentation," 2016 IEEE Conference on Computer Vision and Pattern Recognition (CVPR), Las Vegas, NV, 2016, pp. 2487-2496, doi: 10.1109/CVPR.2016.273.

\bibitem{21} Chen, H., et al.,"DCAN: Deep contour-aware networks for object instance segmentation from histology images."Med Image Anal,vol. 36,pp. 135-146,2017. 

\bibitem{22} Xu, Y., et al.,"Gland Instance Segmentation Using Deep Multichannel Neural Networks."IEEE Trans Biomed Eng,vol. 64,no. 12,2017. 
         
\bibitem{23}
Gadermayr, M., et al.,"CNN cascades for segmenting whole slide images of the kidney."arXiv preprint arXiv:1708.00251, 2017.

\bibitem{24} BenTaieb, A., J. Kawahara, and G. Hamarneh."Multi-loss convolutional networks for gland analysis in microscopy."in IEEE Int Symp Biomedical Imaging. 2016.

\bibitem{25}
Tsung-Yi Lin, Priya Goyal, Ross Girshick, Kaiming He,
and Piotr Dollar, “Focal loss for dense object detection,”
in The IEEE International Conference on Computer Vision
(ICCV), Oct 2017.

\bibitem{26}
F. Milletari, N. Navab, and S.-A. Ahmadi."V-net: Fully convolutional
neural networks for volumetric medical image segmentation."
In 2016 Fourth International Conference on 3D
Vision (3DV), pp. 565–571. IEEE, 2016.

\bibitem{27} Wahab, N., A. Khan, and Y.S. Lee,"Two-phase deep convolutional neural network for reducing class skewness in histopathological images based breast cancer detection."Comput Biol Med,vol. 85,pp. 86-97,2017. 

\bibitem{28} Bulten, W., Pinckaers, H., van Boven, H., Vink, R., de Bel, T., van Ginneken, B., van der Laak, J., de Kaa, C.H.v., Litjens, G.,"Automated gleason grading of prostate biopsies using deep learning." arXiv preprint arXiv:1907.07980,2019.

\bibitem{29} Xing, F., Cornish, T.C., Bennett, T., Ghosh, D., Yang, L.,"Pixel-to-pixel learning with weak supervision for single-stage nucleus recognition in Ki-67 images." IEEE Transactions on Biomedical Engineering, vol.66, pp.3088–3097, 2019. 

\bibitem{30}
de Bel, T., Hermsen, M., Smeets, B., Hilbrands, L., van der Laak, J., Litjens, G., " Automatic segmentation of histopathological slides of renal tissue using deep learning." in: Medical Imaging 2018: Digital Pathology, pp. 1058112, 2018.

\bibitem{31} Gurcan, M.N., et al."Deep learning for tissue microarray image-based outcome prediction in patients with colorectal cancer."in SPIE Medical Imaging. 2016.

\bibitem{32} Szegedy, C., et al. (2014)"Going Deeper with Convolutions." ArXiv e-prints 1409.

\bibitem{33} Krizhevsky, A., I. Sutskever, and G.E. Hinton,"ImageNet classification with deep convolutional neural networks."Communications of the ACM, vol.60, no. 6,pp. 84-90,2017. 

\bibitem{34}
Simonyan K , Zisserman A ."Very Deep Convolutional Networks for Large-Scale Image Recognition."Computer Science, 2014.

\bibitem{35} He, K., et al. (2015)"Deep Residual Learning for Image Recognition."ArXiv e-prints 1512.

\bibitem{36} Wang, D., et al. (2016)"Deep Learning for Identifying Metastatic Breast Cancer."ArXiv e-prints 1606.

\bibitem{37} A. J. Schaumberg, M. A. Rubin, and T. J. Fuchs, “H\&amp;E-stained Whole Slide Image Deep Learning Predicts SPOP Mutation State in Prostate Cancer,” bioRxiv, pp. 064279, 2018.

\bibitem{38}
Ehteshami Bejnordi B, Veta M, Johannes van Diest P, et al."Diagnostic Assessment of Deep Learning Algorithms for Detection of Lymph Node Metastases in Women With Breast Cancer."JAMA, vol.318, no.22,pp.2199–2210,2017.

\bibitem{39} A. Teramoto, T. Tsukamoto, Y. Kiriyama, and H. Fujita, “Automated Classification of Lung Cancer Types from Cytological Images Using Deep Convolutional Neural Networks,” BioMed Research International, vol. 2017, pp. 6, 2017.

\bibitem{40}
Yu K H , Zhang C , Berry G J , et al."Predicting non-small cell lung cancer prognosis by fully automated microscopic pathology image features."Nature Communications,vol. 7,pp. 12474,2016.

\bibitem{41}
Long, J., Shelhamer, E., Darrell, T."Fully convolutional networks for
semantic segmentation."In The IEEE Conference on Computer Vision
and Pattern Recognition (CVPR) (June 2015)

\bibitem{42}                                                          
 Gao Huang, Zhuang Liu, Laurens van der Maaten, and
 Kilian Q. Weinberger, “Densely connected convolutional
 networks,” 2017 IEEE Conference on Computer Vision
 and Pattern Recognition (CVPR),pp. 2261–2269, 2017
 
\bibitem{43}
Chen, L.C., et al.,"DeepLab: Semantic Image
Segmentation with Deep Convolutional Nets, Atrous
Convolution, and Fully Connected CRFs."IEEE
Transactions on Pattern Analysis \& Machine Intelligence,vol, 40,no. 4,pp. 834-848,2018.

\bibitem{44}
Teichmann, Marvin T. T. and Cipolla, Roberto,"Convolutional CRFs
for Semantic Segmentation."arXiv:1805.04777, 2018

\bibitem{45}
 AI Explore platform for real time whole slide segmentation,
http://aiexploredb.ntust.edu.tw/

\bibitem{46}
Li, H., Wang, X. and  Ding, S. "Research and development of neural network ensembles: a survey." Artif Intell Rev, vol. 49,pp. 455–479 (2018).

\bibitem{47}
R. Evans, J. Jumper, J. Kirkpatrick, L. Sifre, T.F.G. Green, C. Qin,
A. Zidek, A. Nelson, A. Bridgland, H. Penedones, S. Petersen, K.
Simonyan, S. Crossan, D.T. Jones, D. Silver, K. Kavukcuoglu, D.
Hassabis, A.W. Senior, “De novo structure prediction with deeplearning based scoring”, In Thirteenth Critical Assessment of
Techniques for Protein Structure Prediction (Abstracts) 1-4
December 2018.

\bibitem{48}
Ronneberger, O., Fischer, P., Brox, T.: U-Net: convolutional networks for biomedical image segmentation. In: Navab, N., Hornegger, J., Wells, W.M., Frangi, A.F.(eds.) MICCAI 2015. LNCS, vol. 9351, pp. 234–241. Springer, Cham (2015).

\bibitem{49}
B. Han et al., “Co-teaching: Robust Training of Deep Neural
Networks with Extremely Noisy Labels,” NeurIPS, pp. 1–
11, 2018.

\bibitem{50}
Tapas Kanungo, David M. Mount, Nathan S. Netanyahu, Christine D. Piatko, Ruth Silverman, and Angela Y. Wu, "An efficient k-means clustering algorithm: Analysis and implementation," IEEE Trans. Pattern Anal. Mach. Intell., vol. 24, pp. 881-892, 2002.



\end{thebibliography}
\end{document}